\documentclass{article}
\usepackage[utf8]{inputenc}
\usepackage{amsmath,amssymb,amsfonts}
\usepackage[left=2cm,right=3cm]{geometry}
\usepackage[linesnumbered,ruled]{algorithm2e}
\usepackage{graphicx}
  \usepackage{authblk}
\newcommand\blfootnote[1]{%
  \begingroup
  \renewcommand\thefootnote{}\footnote{#1}%
  \addtocounter{footnote}{-1}%
  \endgroup
}

\usepackage{subfig}
\usepackage{float}
\usepackage{xcolor}
\newtheorem{definition}{Definition}

\newtheorem{theorem}{Theorem}

\newtheorem{lemma}{Lemma}
\newenvironment{proc}[1][htb]
  {
   \begin{algorithm}[#1]%
  }{\end{algorithm}}

\begin{document} 

\title{Design of Anonymous Endorsement System in Hyperledger Fabric}

\author{Subhra Mazumdar}
\author{Sushmita Ruj}

\affil{Cryptology and Security Research Unit \\ Indian Statistical Institute \\ 203 Barrackpore Trunk Road, Kolkata  700 108, India}
\affil{Email : subhra.mazumdar1993@gmail.com, sush@isical.ac.in }

  \maketitle

%

\begin{abstract}
{Permissioned Blockchain has become quite popular with enterprises forming consortium since it 
prioritizes trust over privacy. One of the popular platforms for distributed ledger solution, \textit{Hyperledger Fabric}, requires a transaction to be \textit{endorsed} or approved by a group of special members known as endorsers before undergoing validation. To endorse a transaction, an endorser mentions its identity along with the
signature so that it can be verified later. However, for certain transactions, difference in opinion may exist among endorsers. Disclosing the identity of an endorser may lead to conflict within the consortium. In such cases, an endorsement policy which not only allows an endorser to support a transaction discreetly, but at the same time takes into account the decision of the majority is preferred. Thus we propose an Anonymous Endorsement System which uses a threshold endorsement policy in order to address the issue. \textcolor{black}{To realize a \textit{t-out-of-n} endorsement policy, using any of the existing threshold ring signature for our endorsement system would have violated the privacy of endorsers as either the identity or the secret key of the endorsers get revealed to the party who recombines the signature after collecting each signature share. All these factors motivated us to design} a new ring signature scheme, called \textit{Fabric's Constant-Sized Linkable Ring Signature} (FCsLRS) with \textit{Transaction-Oriented} linkability \textcolor{black}{for hiding identity of the endorsers}. We have implemented the signature scheme in Golang and analyzed its security and performance by varying the RSA \textcolor{black}{(Rivest-Shamir-Adleman)} modulus size. Feasibility of implementation is supported by experimental analysis. Signature and \textcolor{black}{tag generation time is quite fast and remains constant irrespective of change in message length or endorsement set size for a given RSA modulus value, assuming all the endorsers generates their signature in parallel. Each verifier is required to count and check individual valid ring signature. If the aggregate is above the threshold value, stated by the endorsement policy, then it confirms that the transaction is valid. This increases the verification time depending on the threshold value, but has very little effect on the scalability since generally $t<<n$.} Lastly, we also discuss the integration of the scheme on v1.2 Hyperledger Fabric. }
\end{abstract}

\blfootnote{This work got accepted for publication in \textit{IEEE Transactions on Emerging Topics in Computing, Manuscript Type: Technical Track (Regular Paper)}, DOI (identifier) 10.1109/TETC.2019.2920719 }
\section{Introduction}
In permissionless blockchains, the miners and other participants of the network can stay \textcolor{black}{pseudonymous \cite{biryukov2014deanonymisation},\cite{barcelo2014user}}. The trust is decentralised with no single authority having full control over the functioning of the system\textcolor{black}{\cite{nakamoto2008bitcoin}}. However for political or business organizations, transparency is of utmost importance for fair governance, hence anonymity of participants is not desired. Permissioned blockchain on the other hand offers a closed ecosystem where participants needs permission to join the network and their identity is publicly revealed. This sort of a blockchain is suitable for handling use cases of business network, guaranteeing transparency\textcolor{black}{\cite{davidson2016economics}} and scalability\textcolor{black}{\cite{vukolic2017rethinking}}. Economic incentives, code quality, code changes, and power allocation among peers for such networks are based on the business dynamics. 

\begin{figure}
\centering
  \includegraphics[scale=0.6]{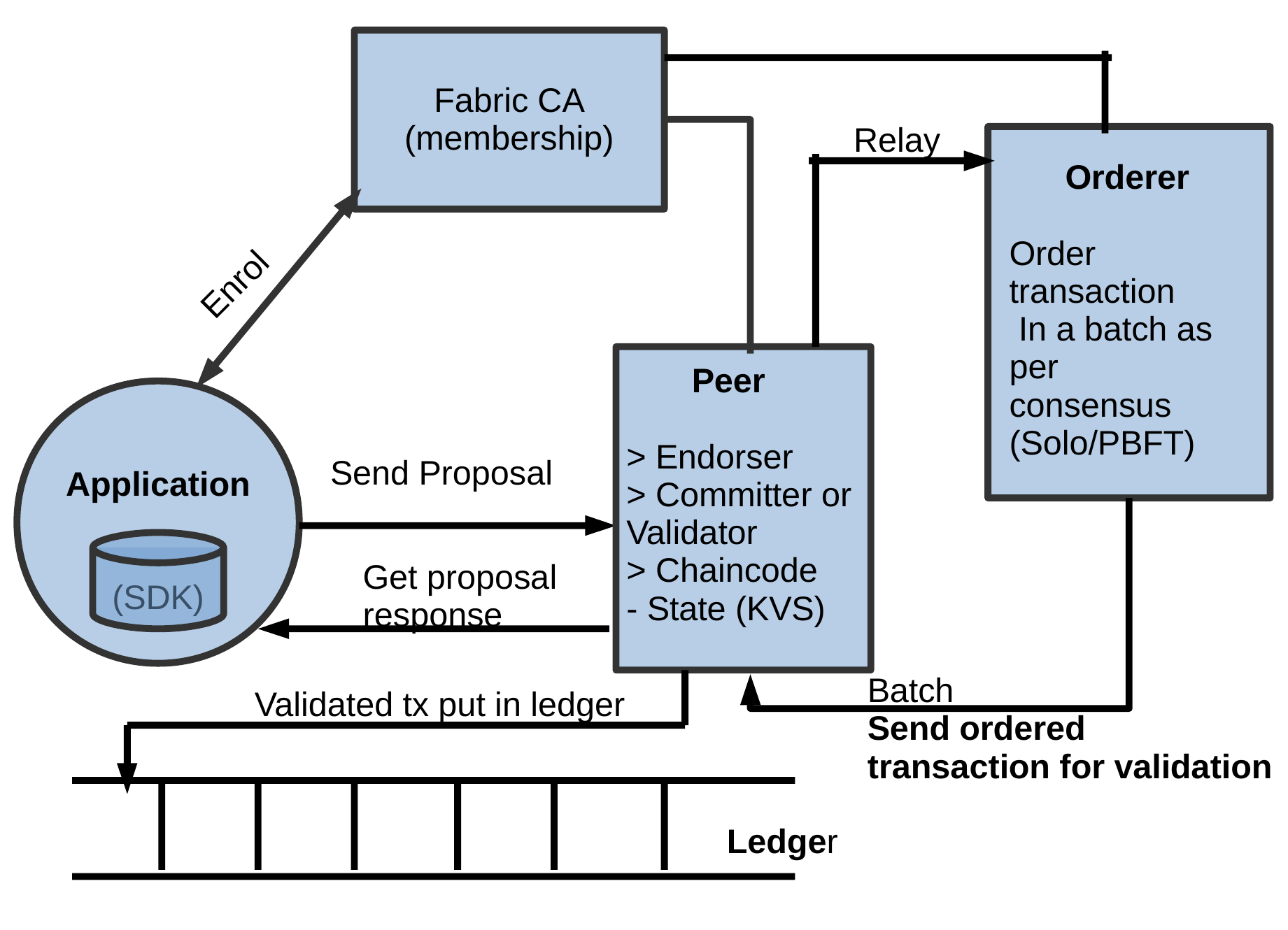}
  \caption{Hyperledger Fabric Architecture}
  \label{fig:imgnode}  
\end{figure}

Hyperledger Fabric is one of the most popular open-source permissioned blockchain frameworks\textcolor{black}{\cite{androulaki2018hyperledger}}. It is quite scalable and robust, hence used mainly for enterprise purpose. Its architecture is illustrated in Fig. \ref{fig:imgnode} \textcolor{black}{(inspired from the diagram in \cite{archi})}. A \textit{Membership Service Provider (MSP)} is responsible for maintaining the identity of all the participants in the system - \textit{clients}, \textit{orderer} or \textit{ordering service} and \textit{validator or peer}\textcolor{black}{\cite{Hyperledger}}. It issues credentials in the form of cryptographic certificates which are used for the purpose of authentication and
authorization. \textit{\textcolor{black}{Fabric CA (Certification Authority)}} acts as a private root CA provider capable of generating keys and certificates needed to configure an MSP\cite{HyperledgerCA}. \textcolor{black}{In Fig. \ref{fig:imgnode}, Fabric CA is the default implementation of MSP interface.} 

\textcolor{black}{We define each network entity briefly given in Fig. \ref{fig:imgnode} \cite{Hyperledger} - a \textit{Client}  acts on behalf of an end-user. It must connect to a peer (of its choice or pre-defined) for communicating with the blockchain. \textit{Peers} are important entities of the network
because they host ledgers and smart contracts. A peer can be a member of multiple channels, and therefore maintain multiple ledgers. The \textit{Ordering Service} or \textit{Orderers} establishes the total order of all transactions in Fabric by reaching a consensus among all peer nodes. Blocks are disseminated to all the peer nodes via peer-to-peer gossip service.} Shared communication channel is formed by members of specific organization, anchor peers acting as leader node for each organization, shared ledger, chaincode application/s and orderers. It allows these network entities to communicate securely. Any entity outside the channel cannot view any activity or status of any member inside a channel.  Data cannot be passed from one channel to another \textcolor{black}{\cite{Hyperledger}}. 

\subsection{Hyperledger Fabric}
\textcolor{black}{Up to version 0.6, Fabric used to follow the \textit{order-execute}   architecture which had several limitations. From version 1.0 onwards, it has been revamped to \textit{execute-order-validate}, ensuring resiliency, flexibility, scalability and confidentiality.  A distributed application for Fabric consists of two parts\cite{androulaki2018hyperledger}:
\begin{itemize}
\item Set of logic encoding the rules for execution of transaction known as \textit{smart contract} or \textit{chaincode}. It needs to be installed across the channel's peer nodes and then instantiated in the channel itself by an authenticated member of the network. 
\item Conditions formed by basic operations of Boolean Algebra - ``AND'' or ``OR'' is used to define which endorser/s must endorse a transaction. This is called as \textit{Endorsement Policy}. The policies can also generalize to logical expressions on sets, such as \textit{t-out-of-n}.
\end{itemize}}

\begin{figure}

\centering
 \includegraphics[scale=0.45]{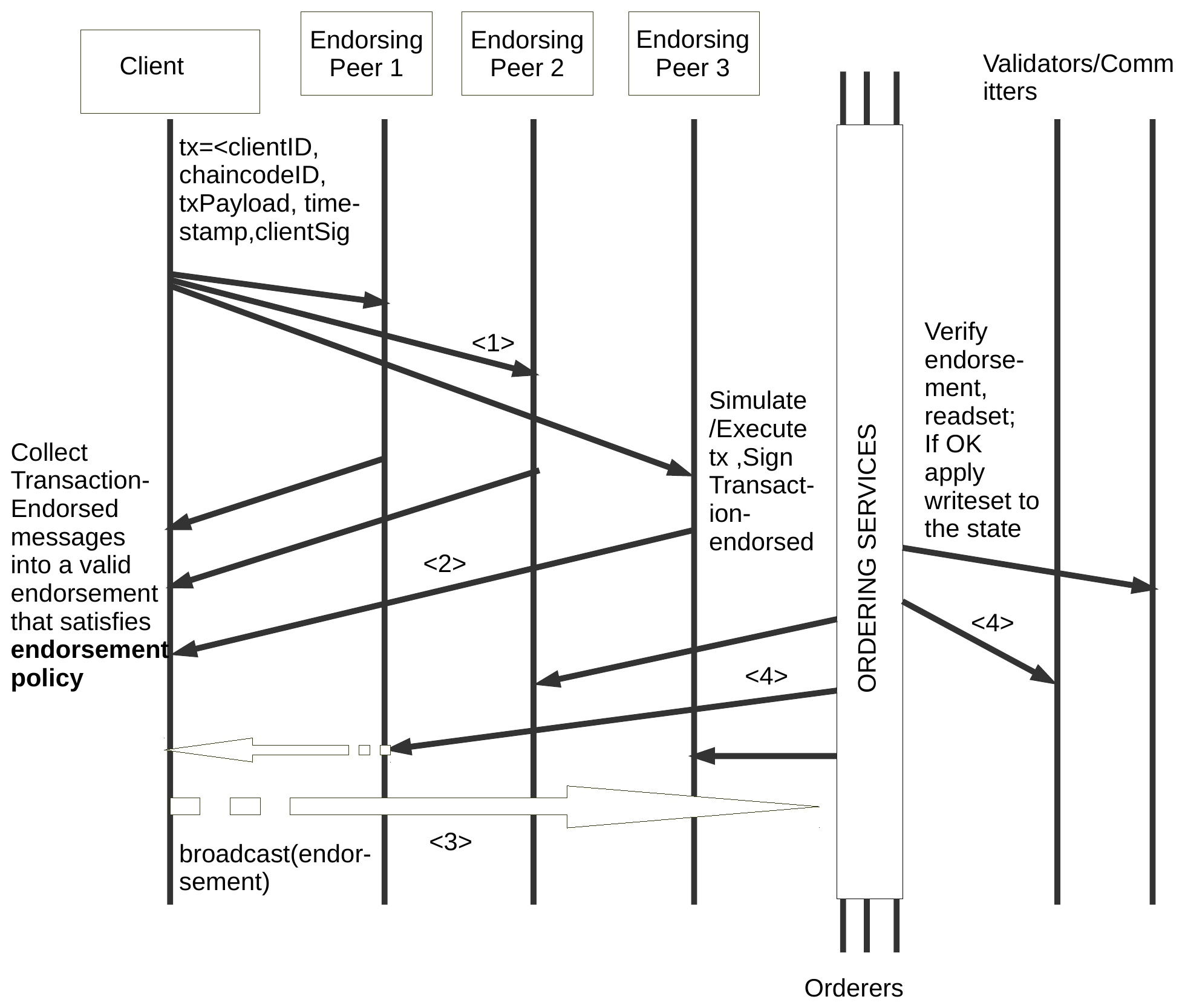}
 
 \caption{Transaction flow in Hyperledger Fabric v1.1}
 \label{imghf}
\end{figure}
\subsubsection{Transaction Flow}
\textcolor{black}{The transaction flow in Fabric involves three main phases - execution, ordering  and validation \cite{androulaki2018hyperledger}. A client comes with transaction request and transmits it to the peer nodes which acts as endorsers. These peer nodes check the validity of the client's signature and checks the compliance of the transaction with the endorsement logic. If all these conditions gets satisfied, the peer \textit{endorses} the transaction by executing it based on the version of keys for the values used in the transaction. This is called the \textit{execution phase} as shown in step $\langle 1 \rangle$ \textcolor{black}{and} step $\langle 2 \rangle$ of Fig. \ref{imghf}. \textcolor{black}{Transactions are executed in parallel by the endorsers to increase the throughput.} The executed transaction along with the endorser's signature is sent back to the client node. If the endorsement policy is satisfied, it is broadcasted to the orderers nodes. In the \textit{ordering phase} (step $\langle 3 \rangle$ of Fig. \ref{imghf}), a pluggable consensus protocol is used to get a total ordering on the sequence of endorsed transactions grouped into blocks. These are broadcasted to all peer nodes for \textit{validation} (step $\langle 4 \rangle$ of Fig. \ref{imghf}). All peers deterministically validate the transactions in the same order by checking satisfaction of endorsement policy (inspect the identity and signature of each endorser) and version number of the keys present in the local key-value store. \textcolor{black}{ Checking of the version number ensures that the data that was read during chaincode execution did not change during endorsement. If it remains same then the transaction is marked valid and it gets committed to the ledger state database. If not, then the transaction is considered as invalid and is not applied to the ledger state database. The client gets information about the status of the transaction. For invalid transaction, it can either apply error handler, retry it or abort it for the time being.} Note that any interaction among nodes occurs through messages send via the channels and are authenticated via digital signatures. }
\subsubsection{Limitation of the existing architecture}
Revealing the identity of an endorser to the peer nodes may not be suitable for sensitive transactions\textcolor{black}{\cite{cambodia},\cite{chile}. Consider a referendum on allocation of funds by World Bank to a developing country, sharing hostile relation with influential developed countries. The referendum succeeds if majority of them vote in favor of the motion. However the voting process is biased. Donor countries, though less in number, have more influence on any decision taken than any borrower country. Economically weaker countries have no role to play. If we deploy the voting process on Hyperledger Fabric with countries forming the set of endorsers, the result will be biased in favour of donor countries. Over here, the setup of Hyperledger Fabric fails to ensure a fair judgement as it reveals the identity of the country involved in endorsement. }

Blockchain was created to solve the specific problem of making a process or execution trustworthy even when all participants remain anonymous. In permissionless setting, anyone can join the network without credentials\textcolor{black}{\cite{nakamoto2008bitcoin}}. On the contrary, in the permissioned setting, a participant's credential is important and must be \textcolor{black}{known} by everyone within the network\textcolor{black}{\cite{vukolic2017rethinking}}. However, keeping in mind the requirement of certain use cases, Hyperledger Fabric intends to integrate \textit{Identity Mixer MSP (Membership Service Provider)} \cite{camenisch2002design} in its future releases where client and/or peer nodes can sign the transaction by remaining anonymous, generating unlinkable signature. With the aid of zero-knowledge proofs\textcolor{black}{\cite{feige1988zero}}, a peer node can prove the correctness of the generated signature, without the verifier obtaining extra information. A privacy preserving distributed ledger, \texttt{zkLedger} \cite{narula17zkledger} claims to support strong transaction as well as participants privacy (for hiding transaction value it uses Pedersen commitments) but at the same time ensures fast and provably correct auditing, with the help of Schnorr-type non-interactive zero-knowledge proofs. But none of them have put their focus on hiding the identity of endorser. This motivated us to design of a bias-free \textit{Anonymous Endorsement System}, allowing provision of an endorsement policy which reveals just the total count of the vote a referendum or a policy proposed (termed as transaction in Fabric) by the client has received from the members of a pre-specified endorsement set without explicitly specifying who has endorsed. 
%

A threshold endorsement policy requires at least \textit{t-out-of-n} endorsers to 
approve the transaction without explicitly mentioning their identity. Any implementation of threshold cryptosystem might seem an obvious answer, but it does not ensure anonymity of endorser \textcolor{black}{\cite{Bresson}} (explained later in Section \ref{rw}). Instead, we use \textit{Ring Signature} based on \textit{Signature of Knowledge}\textcolor{black}{\cite{chase2006signatures}} (Definition \ref{sok}). All the endorsers of the channel form the group members and it is computationally infeasible to determine which endorser has produced the signature. Detaching the identity of signer from the signature can lead to the problem of double-signing whereby a member tries to endorse a transaction more than once. We prevent this by making multiple signatures generated by a particular signer for a particular session linkable \textcolor{black}{\cite{au2006short},\cite{tsang2005short}}. Verifiers simply count each valid endorsement in order to check whether a threshold endorsement policy is satisfied.


\subsection*{Our Contributions:}
This paper makes the following contributions\textcolor{black}{:}
\begin{itemize}
\item We have proposed an \textit{Anonymous Endorsement System} which implements a simple threshold endorsement policy which requires at least $t$ out of $n$ ($1\leq \ t\ \textcolor{black}{\leq} \ n$) endorsers to approve a transaction without explicitly mentioning their identity.
\item We have given the construction of a constant sized linkable ring signature, Fabric's Constant-Sized Linkable Ring Signature (\textit{FCsLRS}), to hide the identity of an endorser.
\item A new linking criterion, called as \textit{``transaction-oriented''} linkability, is used, which prevents an endorser from signing the same transaction more than once.
\item  We have implemented this scheme using Go Programming Language or \textit{Go}, version 1.10 \cite{golang} and analyzed its performance. \textcolor{black}{Golang based smart contracts are fully supported by Hyperledger Fabric \cite{Hyperledger}.} The code for \textit{FCsLRS} is available in \cite{Code}.
\item A detailed description of the integration of the scheme on v1.2 Hyperledger Fabric has been discussed.   
\end{itemize}
\subsection{Organization of the paper}
\texttt{Section \ref{rw}} states the related works on various signature schemes and their application in blockchain system. Mathematical notations and basic definitions have been discussed in \texttt{Section \ref{math}}. In \texttt{Section \ref{AESys}}, we describe our proposed anonymous endorsement system. Construction of Fabric's Constant-Sized Linkable Ring Signature (\textit{FCsLRS}) scheme is stated in \texttt{Section \ref{RS}}. \texttt{Section \ref{Chap:security}} describes the security model. Performance Analysis is given in \texttt{Section \ref{Chap:performance}}.
\texttt{Section \ref{integrate}} elucidates, in detail, the steps needed to integrate the \textit{FCsLRS} scheme with Fabric. \texttt{Section \ref{integrate}} elucidates, in details, the integration of \textit{FCsLRS} scheme with Fabric. Finally the paper is concluded in \texttt{Section \ref{conc}}.

\section{Related Work}
\label{rw}
Even though hiding identity using ring signature is a quite well studied area \cite{camenisch1999separability},\cite{camenisch1997efficient},\cite{rivest2001leak},\cite{cramer1994proofs}, its applicability in cryptocurrencies and blockchain is being recently explored. 

Monero, an anonymous cryptocurrency, improves on its existing \textit{Cryptonote} \cite{van2013cryptonote} protocol by using
a new efficient Ring Confidential Transactions protocol - \texttt{RingCT 2.0} \cite{sun2017ringct}.\textit{ ``Over here, the size of signature is independent of the number of groups of input accounts in a transaction. It makes use of the following cryptographic primitives - Pedersen commitment, accumulator with one-way domain and signature of knowledge related to the accumulator\cite{sun2017ringct}.''} \textcolor{black}{However the protocol used for anonymizing sender and receiver identity and also the transaction value, cannot be used for voting/endorsement policy.}. \textit{ZCash} requires a trusted set up stage, but after that the system is entirely anonymous, making use of zero-knowledge proof for verification \cite{hopwood2016zcash}. \textcolor{black}{But this violates the motivation of our scheme where we do not want to depend on any third party, even during setup phase}. Hardjono et al. \cite{hardjono2014anonymous} presented a new architecture called \textit{ChainAnchor}, address the issue of retaining user anonymity, introducing the concept of \textit{semi-permissioned} blockchains \textcolor{black}{but it makes use of TPM tamper resistant hardware, which is not in the purview of our proposed system.}

\textcolor{black}{Existing ring signature schemes like Unique Ring Signature \cite{zhang}, Yuen et al. \cite{yuen2012efficient} have signature size proportional to the size of ring. Threshold signature in Bresson et al. \cite{Bresson} does not ensure full anonymity. Over here, the signature combiner (any trusted third party) needs to know the id and secret key of the signers in order to combine the signature. \textit{t-out-of-n} signature scheme by Tsang et al.\cite{tsang2004separable} allows event-oriented linkability which can result in double signing for our use case.} Some are too complex to be implemented with signature size being dependent on the ring size (as in Yuen et al. \cite{yuen2012efficient}). \textcolor{black}{A detailed comparative analysis of these signature schemes has been provided as supplemental material (labeled as Table 1) in tabular format.} Several threshold signature schemes for enhancing Bitcoin security has been proposed in Goldfeder et al. \cite{goldfeder2015securing}, Gennaro et al. \cite{gennaro2016threshold} and Kogias et al. \cite{kogias2016enhancing} \textcolor{black}{but these signature scheme suffers from the same problem of trusting the signature combiner as was stated before.}  

To this day, the only work which has focused on threshold signatures in Hyperledger Fabric is \cite{stathakopoulou2017threshold}. They have identified numerous potential application of threshold signature which can be used for group of Certificate Authorities, Byzantine Consensus protocols, chaincode applications and transaction validation. For this they have compared the performance of threshold signature schemes - threshold RSA \textcolor{black}{(Rivest-Shamir-Adleman)} signature/threshold BLS \textcolor{black}{((Boneh-Lynn-Sacham)} signature, against multisignature. In such schemes, a trusted third party is requested to generate the key shares and distribute them among signing parties (\textit{Kate et al. \cite{kate2010distributed}} proposed a scheme for
distributed key generation scheme but it comes with fair amount of
computation overhead). Any entity performing the task of signature combination in threshold signature scheme must know id value of the signer so that it can compute the Lagrange coefficient.  Verifier verifies just one signature instead of verifying each signature submitted by the endorsers. However, it leaks the identity of endorser to the entity who performs the task of recombining signature shares. 

\textcolor{black}{Our proposed anonymous \textit{t-out-of-n} endorsement system allows a set of endorsers to endorse any transaction without revealing the identity of any of the endorsers to any member within the network, including the other members of endorsement set. But since $t$ such signatures are generated, a verifier needs to verify it individually one at a time, unlike threshold signature where verifier needs to verify just one signature. This increases the amount of verification time in total, although the time taken to verify one signature by an endorser is constant, independent of the size of the ring.} Since our problem statement demands anonymity on the identity of endorsers and not efficient transaction validation, we consider use of linkable ring signature for our anonymous endorsement system \cite{mazumdar2018thesis}. It avoids all the complexities associated with the implementation of threshold cryptosystem.

\footnotetext[1]{Exponentiation operation is of the form $g^a$ for base $g$, multibase exponentiation operation is of the form $g^a.h^b$ for base $g,h$.}

 \begin{table*}[h]
  \centering

 \flushright  \caption{Comparison of Ring Signature Schemes}
   \label{tab:appendix}


  \scalebox{0.7}[0.6]{
   \hspace{-1cm}

  \begin{tabular}{|c c c  c c c c|} 
 \hline
  Scheme & Signature  & Security  & Linking  & Signing \footnotemark[1]
  &Verify \footnotemark[1] &Problem \\
  &Size &Notions  &Complexity &Complexity &Complexity &Encountered \\ 
   \hline
 \textit{1-out-of-n}  & $\mathcal{O}$(1)& Unforgeability,  & $\mathcal{O}$(1): check   &Uses Signature based on     & 7 multibase &  \\ 
Ho. Au \textit{et al.}  \cite{au2006short}  & & Linkable Anonymity,       &linkability tag & Proof of Knowledge, (n+2)  & exponentiation & Adversary can \\
    & & Linkability,      & group oriented & exponentiation and  &   & corrupt the  \\
      & & Non slanderability     &linkability  & 7 multibase &  &signer $\mathcal{S},$\\ 
      & & - all wrt adversarially  & &exponentiation & & overhead of certificate\\ 
      & &chosen keys  & & & &check, need event \\ 
     &&&&& &oriented linkability since \\ 
      &&&&&& ring members are fixed\\
      
 \hline
 \textit{t-out-of-n}\cite{Bresson} &$\mathcal{O}(l.2^t n\log n)$&Unforgeability,& unlinkable &$t.2^t \log n + t$ symmetric&$t.2^t \log n$ symmetric&\\ 
Bresson \textit{et al.} &&\textit{t-CMA} secure&&cipher operation, &cipher op,&Prover may \\ 
&&anonymity&&$n.2^t \log n + n$&$t.2^t \log n$ exponen-&be malicious,\\ 
&&&&exponentiation&tiation&t signers need\\ 
&&&&&&to share their\\ 
&&&&&&secret keys, compu-\\ 
&&&&&&tationally expensive,\\ 
&&&&&&double signing can't\\
&&&&&&be prevented\\ 
 \hline
\textit{t-out-of-n} &$\mathcal{O}$(n) &Unforgeability,  & $\mathcal{O}(n^2)$ &2(n+d) exponentiations &$\mathcal{O}$(n) multibase & \\
  Tsang \textit{et al.}\cite{tsang2004separable} &&Linkable Anonymity, &Event-oriented & and 2(n-d) & exponentiations &Prover/Signer may  \\
  &&Event-oriented Linkability, &Linkability &multibase & &create 2 different \\
  &&Non slanderability. &&exponentiation &&event-id (double\\ 
    && && & &signing possible),\\ 
       && && &&Problem of CDS \cite{cramer1994proofs}\\
       && && & &scheme exists,\\
&& && &&sharing of secret key   \\
      
&& && && with prover $\mathcal{P}$, when\\
&& && &&  $\mathcal{P}$ gets compro-\\
&& && && mised, is not desired \\
  \hline
  \textit{t-out-of-n}\cite{yuen2012efficient}&$\mathcal{O}(t.\sqrt{n})$&Unforgeability,&$\mathcal{O}(t \log t)$, event-&(8t+4t$\sqrt{n}$) exponentiation, &(8t+8t$\sqrt{n}$) pairing,&More complex than  \\ 
  Yuen \textit{et al.}&&linkable anonymity &oriented linka- &(4t+2t$\sqrt{n}$) multibase&2t exponentiation,&\textit{1-out-of-n} signa-\\ 
&&event-oriented link- &bility&exponentiation&t one-time veri-&ture scheme\\ 
&&ability, non- &&&fication &, dependence on \\ 
&&slanderability &&& &event id\\ 
  \hline
\textit{URS}  &$\mathcal{O}$(n)&Unforgeability,&$\mathcal{O}$(1) - tag is &2n-1 multibase& 2n multibase &Computationally expensive \\ 
Franklin,Zhang\cite{zhang},  &&secure linkability, &hash of message,&exponentiation and &exponentiation&scheme, not \\ 
\cite{rebekah}&& and restricted &ring members and& 1 exponentiation &&yet extended to\\ 
&&anonymity &private key && &\textit{t-out-of-n}\\ 
&&&of signer&& &scheme.\\ 
  \hline
Our proposed &$\mathcal{O}(1),$ &Unforgeability &$\mathcal{O}(1)$, & 11 exponentiations &10 multibase &Extension to \\
signature scheme &constant size &Linkability, &transaction ori- &and 5 multibase &exponentiations &\textit{t-out-of-n} \\
\textit{FCsLRS} & &Linkable Anonymity,&ented linkability &exponentiations &and 6 exponentiation &signature scheme \\
& &Non-slanderability, & & & &not efficient.\\
& &-all wrt adversarially chosen keys & & & &\\
\hline
\end{tabular}
}

\end{table*}

\section{Preliminaries}
\textcolor{black}{For the proposed endorsement system, we provide a construction of ring signature based on signature of knowledge. It involves proving set of NP statements in zero knowledge \cite{chase2006signatures}. We define some mathematical terms used as well as hardness assumptions needed to prove the security of the scheme.}
\label{math}
\subsection{Mathematical Notations}
We define the mathematical notations which will be used in our construction of FCsLRS scheme :
\begin{itemize}
\item   $\lambda,l,\mu \in \mathbb{N} : \lambda > l - 2, l/2 > \mu + 1$ be the security parameters.
\item \texttt{RSA}$_\lambda$ be the set of RSA integers of size $\lambda$.
\item A number $p$ is a \textit{safe prime} if $p=2p'+1$ and both $p$ and $p'$ are odd primes.
\item A number $N$ is an \textit{RSA integer} if $N=pq$ for distinct safe primes $p$ and $q$ where $p=2p'+1$ and $q=2q'+1$. It is termed as a rigid integer $|p|=|q|$. 
\item Set of $\lambda$-bit rigid integers are denoted by \texttt{Rig}$_\lambda$.
\item $QR(N)$ denotes the group of quadratic residues modulo $N$ of order $p'q'$.
\end{itemize}
\vspace{-0.4cm}
\subsection{Hardness Assumptions}
\begin{itemize}
\item \textbf{Decisional Diffie-Hellman (DDH) Assumption.} \cite{au2006short} 
\textit{``Consider a group $G$ of order $q$, $q$ is prime and let $g$ be generator of $G$. Given $a,b,c \in_R \mathbb{Z}_q$, there exists no probabilistic polynomial time (PPT) algorithm that can distinguish two distributions $\langle g,g^a,g^b,g^{ab}\rangle$ and $\langle g,g^a,g^b,g^c\rangle$ with non-negligible probability over 1/2 in time polynomial in $q$.''}
\item \textbf{Strong RSA (SRSA) Assumption.} \cite{au2006short} \textit{``Given input a random RSA integer $N$ and a value $z \in_{R} QR(N)$, there exists no probabilistic polynomial time (PPT) algorithm which can return $u \in \mathbb{Z}_N^*$ and $e \in \mathbb{N}$ such that $e > 1$  and $u^e = z(mod \ N )$, with non-negligible probability and in time polynomial in $\lambda$.''}
\item \textbf{Link Decisional RSA (LD-RSA) Assumption.}\cite{au2006short} \textit{``Given input a random RSA integer $N$, $\hat{g} \in_{R} QR(N), n_0=p_0q_0$ and $n_1=p_1q_1$ where $p_0,q_0,p_1,q_1$ are sufficiently large random primes of size polynomial in $\lambda, \hat{g}^{p_b+q_b}$ where $b \in_R \{0,1\}$, there exists no PPT algorithm which returns $b'=b$ with probability non-negligibly over 1/2 and in time polynomial in $\lambda$.''}
\end{itemize}

\subsection{Building Blocks}
\textcolor{black}{For construction of Signature of Knowledge, we need help of sigma protocol which is defined below.}

\begin{definition}
(\textbf{$\Sigma$-Protocols}.) \cite{dodis2004anonymous} \textit{It is an efficient 3-round two-party protocol defined over an \textbf{NP}-relation $R$. For every input $(x,secret)$  given to prover $P$ and $x$ given to verifier $V$, the first round, initiated by prover $P$, yields a commitment message \texttt{COM}. In the second round, verifier $V$ replies with a random \textit{challenge} message \texttt{CH}. The last round by $P$ concludes by sending \textit{response} message \texttt{RES}. Finally, an honest verifier will output a 0 or 1, provided the transcript $\pi=$(\texttt{COM,CH,RES}) is valid and prover $P$ possesses the \textit{secret}.}
\end{definition}
A $\Sigma$-protocol satisfies - \textit{Special Soundness} and \textit{Special Honest-Verifier Zero-Knowledge} property. This protocol can be efficiently constructed under the assumption that one way functions are easy to compute but hard to invert.

%
%
%


\vspace{0.5cm}
\textcolor{black}{A machine called knowledge extractor guarantees that the prover actually holds the witness i.e. success probability of knowledge extractor extracting witness is proportional to the success probability of prover convincing the verifier.}
\begin{definition}
\label{def:ke}

(\textbf{Knowledge extractor}). \cite{Goldreich:2006:FCV:1202577} \textit{ On input $x$, auxiliary input $\tilde{z}$ and random input $r$ : $x \in L_R$, \textcolor{black}{where $L_R$ is the \textbf{NP-language} to $R$ defined as $L_R=\{x|(\exists secret)[(x,secret)\in R]\}$} and  $\tilde{z},r \in \{0,1\}^*$, let V outputs 1, after interacting with prover specified by $P_{x,\tilde{z},r}$ with probability $p(x,\tilde{z},r)$ and $\kappa(.)$ be  error, $\kappa : \mathbb{N} \rightarrow [0,1]$.
A probabilistic oracle machine $K$ is called a \textbf{(universal) knowledge extractor}, if on input x (same as that given to V) and access to oracle $P_{x,\tilde{z},r}$, it outputs a solution $y \in R(x)$ within an expected number of steps bounded by
\begin{equation*}
\frac{q(|x|)}{p(x,\tilde{z},r)-\kappa(|x|)}
\end{equation*}
where q(.) is a positive polynomial, provided $p(x,\tilde{z},r) > \kappa(|x|)$.}


\end{definition}

%
%
\vspace{0.5cm}
\textcolor{black}{Sphere Truncation of a group enables to prove that it is equally hard to break the discrete logarithm hardness in spite of significant reduction in the size of group. }
\begin{definition}
\textbf{(Sphere Truncations of Quadratic Residues).}\cite{dodis2004anonymous}
\label{ST} \textit{Given that $N$ is a RSA integer where $N = pq$, we define a sphere denoted by $S(2^l , 2^\mu ) = \{2^l - 2^\mu + 1,\ldots , 2^l + 2^\mu - 1 \}$  for two parameters $l, \mu \in N$ where  $|S(2^l , 2^\mu )| = 2^{\mu+1} - 1$. Any random variable $a^x$ with $x \in_{R} S(2^l , 2^\mu )$ is indistinguishable from the uniform distribution over $QR(N)$ provided factoring is hard and sphere $S(2^l,2^\mu)$ is sufficiently large but not of the order of $QR(N)$. In simple terms, this means that a probabilistic polynomial-time observer cannot distinguish whether a value is selected from $S(2^l , 2^\mu)$ or $QR(N)$.}
\end{definition}
\vspace{0.5cm}
\textcolor{black}{ In order to plan complex proofs of knowledge for protocols operating over groups of unknown order in general like for group $QR(N)$, discrete-log relation set are quite useful.}
\begin{definition}
\textbf{(Discrete-log Relation Sets).} \cite{dodis2004anonymous} \textit{For a group $G$ of unknown order, a discrete-log relation set $R$ with $z$ relations over $r$ free variables, $\alpha_1,\ldots,\alpha_r$, and $m$ objects is a set of relations defined over the objects $A_1,\ldots,A_m \in G$, such that 
\begin{itemize}
\item[-] each free variable $\alpha_j$ is assumed to take value in a finite integer range $S(2^l_j,2^\mu_j)$ where $l_j,\mu_j \geq 0$.
\item[-] $i^{th}$ relation in the set $R$ is specified by a tuple $\langle a_1^i,\ldots,a_m^i \rangle \ \textrm{so that each} \ a_j^i$ is selected to be one of the free variables $\{\alpha_1,\ldots,\alpha_r\}$ or an element of $\mathbb{Z}$. The relation is to be interpreted as $\Pi_{j=1}^m A_j^{a_j^i}=1$.
\end{itemize}}
\end{definition}

A \textit{discrete-log} relation set $R$ is said to be \textit{triangular}, if for each relation $i$ involving the free variables $\alpha_1 ,\alpha_2 ,\ldots \alpha_k$,
it holds that the free-variables $\alpha_1 ,\alpha_2 ,\ldots \alpha_k$ are contained in relations $1,\ldots,i-1$.

\vspace{0.5cm}
\begin{definition}
\textbf{(Signature of Knowledge).}
\label{sok}
\textit{Instead of Challenger supplying the challenge value to Prover in  three-round $\Sigma$-protocols or Honest-Verifier-Zero-Knowledge (HVZK) Proof of Knowledge (PoK) protocols, setting the challenge to the hash value of the commitment concatenated with the message to be signed \cite{feige1988zero} by the Prover transforms it into a signature scheme known as Signature based on Proof of Knowledge or simply `Signature of Knowledge (SoK)'\cite{camenisch1997efficient},\cite{chase2006signatures}. Security of this scheme in the random oracle model is defined in\cite{pointcheval1996security},\cite{bellare1993random}.}  
\end{definition}
%
%

\vspace{0.5cm}
\textcolor{black}{Our objective is to design a signature scheme where the size of the endorser's signature  is independent to the number of endorsers present. With the help of accumulator, we can accumulate the public key of all the members in the endorsement set and proof membership of the endorser to that set.}
\begin{definition}
\textbf{(Accumulators with One-Way Domain).} \cite{dodis2004anonymous}\cite{tsang2005short} \textit{An \textit{accumulator family} is a pair $(\{F_\lambda\}_{\lambda \in \mathbb{N}}, \{X_\lambda\}_{\lambda \in \mathbb{N}})$, where $(\{F_\lambda\}_{\lambda \in \mathbb{N}}$ is a sequence of families of functions such that each $f \in F_\lambda$ is defined as $f: U_f \times X_f^{ext} \leftarrow U_f$ for some $X_f^{ext} \subseteq X_\lambda$. It also satisfies the properties - efficient generation (in polynomial time in $\lambda$) and efficient evaluation (in polynomial time in $\lambda$). 
For all $\lambda \in \mathbb{N}$, $f \in F_\lambda$, $u \in U_f, x_1, x_2 \in X_\lambda$,}
\begin{equation}
f(f(u,x_1),x_2)=f(f(u,x_2),x_1)
\end{equation}
\textit{$\{X_\lambda\}_{\lambda \in \mathbb{N}}$ is referred to as the \textit{value domain} of the accumulator. Due to the property of quasi-commutativity, such value is independent of the order of the $x_i's$ and will be denoted by $f(u,X)$. For any $\lambda \in \mathbb{N}, f \in F_\lambda$ and $X = \{x_1,\ldots,x_s\} \subset X_\lambda$, $f(\ldots f(u,x_1),\ldots,x_s)$ is the \textit{accumulated value} of the set $X$ over $u$. }
%

\textit{Based on the hardness assumption of Strong RSA, an \textit{accumulator with one-way domain}\cite{au2006short}  is a quadruple $(\{F_\lambda\}_{\lambda\in \mathbb{N}},\{X_\lambda\}_{\lambda\in \mathbb{N}},\{Z_\lambda\}_{\lambda\in \mathbb{N}},\{R_\lambda\}_{\lambda\in \mathbb{N}})$, such that the pair $(\{F_\lambda\}_{\lambda\in \mathbb{N}},\{X_\lambda\}_{\lambda\in \mathbb{N}})$ is a collision-resistant accumulator, each $R_\lambda$ is a relation over $X_\lambda \times Z_\lambda$ with the following properties:
\textit{(efficient verification)}. There exists an efficient algorithm $D$ that on input $(x,z) \in X_\lambda \times Z_\lambda$, returns 1 if and only if $(x,z) \in R_\lambda$.
\textit{(efficient sampling)}. There exists a probabilistic algorithm $W$ that on input 1$^\lambda$ returns a pair $(x,z) \in X_\lambda \times Z_\lambda$ such that $(x,z) \in R_\lambda$, $z$ is the \textit{pre-image} of $x$. 
\textit{(one-wayness)}. It is computationally hard to compute any pre-image $z'$ of an $x$ that was sampled with $W$. Formally, given a negligible value $\nu(\lambda)$, for any adversary $\mathcal{A}$:}
\begin{equation}
Pr[(x,z)\xleftarrow{R} W(1^\lambda); z' \xleftarrow{R} \mathcal{A}(1^\lambda,x)\ \vert\ (x,z')\in R_\lambda] = \nu(\lambda)
\end{equation}
\textit{For $\lambda \in \mathbb{N}$, the family $F_\lambda$ consists of the exponentiation functions modulo $\lambda$-bit rigid integers :}
\begin{equation}
\label{eqn1}
\begin{matrix}
f : QR(N) \times \mathbb{Z}_{N/4} \rightarrow QR(N)\\
f : (u,x) \rightarrow u^x \mod N
\end{matrix}
\end{equation}
where $N \in$ \texttt{Rig}$_\lambda$.

\textit{The accumulator domain $\{X_\lambda\}_{\lambda \in \mathbb{N}}$ is defined by:}
\begin{equation}
X_\lambda = \{ e \ \textrm{prime} \ | \ (\frac{e-1}{2} \in \texttt{RSA}_l) \wedge (e \in S(2^l,2^\mu))\}
\end{equation}
\textit{where $S(2^l,2^\mu)$ is the integer range $(2^l-2^\mu,2^l+2^\mu)$ that is embedded within $(0,2^\lambda)$ with $\lambda - 2 > l \ \textrm{and} \ l/2 > \mu + 1$. 
The pre-image domain $\{Z_\lambda\}_{\lambda \in \mathbb{N}}$ and the one-way relation $\{R_\lambda\}_{\lambda \in \mathbb{N}}$ are defined as follows:}
\begin{equation}
Z_\lambda= \Bigg\{
\begin{matrix}
(e_1,e_2)\ | \ e_1,e_2 \ \textrm{are distinct}\ l/2-bit\\ \textrm{ primes and }
e_2 \in S(2^\frac{l}{2},2^\mu) \\
R_\lambda = \{(x,(e_1,e_2))\in X_\lambda \times Z_\lambda \ | \ (x=2e_1e_2+1)\} 
\end{matrix}
\Bigg\}
\end{equation}
\end{definition}

\section{Our Proposed Anonymous Endorsement System }
\label{AESys}
To address the problem of biased endorsement policy as well as ensuring privacy of endorsers, we have designed an anonymous endorsement system. To ensure privacy of the system, we have proposed a new ring signature scheme which is discussed in Section \ref{RS}.
 
In Hyperledger Fabric, membership service provider (MSP) identifies the parties, who are the members of a given organization in the blockchain network. The endorsement set for a particular chaincode is presumed to be predefined and remains fixed for a long time, unless any of the members get revoked. The right measure of ``signature size'' constructed for each transaction must not involve explicit description of the ring members (endorsers for this case). A one-time computation of accumulation of public keys proportional to the size of the ring needs to be performed by the \textit{Fabric CA} and communicated to all the verifiers present in the network. This constant-sized information allows signers to generate or verifiers to verify many subsequent signatures in constant time.

\subsection*{Entities present in the network}
\textit{Fabric CA(Certification Authority) Server} \footnote{It is a private root CA provider capable of managing digital identities of Fabric participants that have the form of X.509 certificates} issuing enrolment certificates to all the peer nodes (endorser and validators). Setup mentioned in \cite{HyperledgerCA}.\\ \textit{Client} : An entity lying outside the blockchain network, having a transaction request. A peer node, within the network, acts as a proxy for the client node.\\ 
\textit{Endorser set} $\mathcal{E}$ : A pre-defined set to be specified before instantiation of chaincode. Members of this set, based on a given endorsement policy, decides on whether to endorse a transaction.\\ \textit{Signer} $S$ : A member of the endorsement set $\mathcal{E}$ which executes the ring signature algorithm on the transaction response packet for the endorsed transaction. \\ \textit{Verifier/Validator set} $\mathcal{V}$ : Validator nodes verify whether the signature was generated by a valid member of the endorsement set.
 \subsection*{Requirement of the signature scheme}
Signature and tag generated must be short. The signature generation and verification must be computationally efficient. None of the entities must get access to any secret of the signer. Apart from this, the following two properties must be ensured for the scheme : 
\begin{itemize}
\item[-]With negligible probability, valid signatures generated according to specification
fails to get \textbf{accepted} during verification.\\ 
\item[-]With negligible probability, two signatures signed according to specification, generated by the same signer on the same transaction for a given set of endorsers
fails to get \textbf{linked}.
\end{itemize}
\subsection{Proposed construction of Fabric's Constant-Sized Linkable Ring Signature (FCsLRS)}
\label{RS}
In this section, we propose a new ring signature scheme  called as \textit{Fabric's Constant-Sized Linkable Ring Signature}(FCsLRS) for a fixed set of endorsers and discuss the construction details. Our construction is inspired by the signature scheme obtained by applying \textit{Fiat-Shamir transformation} to the \texttt{Identification Protocol} suggested in Dodis et al.\cite{dodis2004anonymous}. Previously, this identification protocol has been used as a short signature scheme by Tsang, et al.\cite{tsang2005short} and  Ho Au, et al.\cite{au2006short} for e-Cash, e-voting and attestation. But none of them could have been used directly for our endorsement system, \textcolor{black}{justification made later while describing the algorithm \texttt{Sign}.} \textcolor{black}{A transaction level tag is needed here which prevents an endorser from endorsing a transaction more than once but it can endorse two different transaction. This is not present in any of the previous construction. In \cite{dodis2004anonymous},\cite{au2006short},\cite{tsang2005short}, the public values constructed by prover to prove validity of public key in \textit{Signature of Knowledge} allows an adversary to easily identify the endorser. This is because the public key of all the endorsers is known to all network entities and hence the identity can be revealed in polynomial time.} 

Considering $n$ to be the number of members in the endorsement set and $t$ ($1 \leq t<n$) to be the threshold value. FCsLRS is represented as a tuple (\texttt{Init, KeyGen, AccumulatePubKey, GeneratePubKeyWitness, Sign, Verify, Link}) of seven polynomial time algorithms, which are described below. 
\begin{itemize}
\item[-] \textbf{\texttt{Init}}. On input of security parameters, \textit{Fabric CA (Certificate Authority)} prepares a collision-resistant accumulator with one-way domain. A generator $u \in QR(N)$ is picked up uniformly at random, where $N \in$ \texttt{Rig}$_\lambda$. 
Public parameters $g,h,y,t,s,\zeta \in QR(N)$, is also generated. These parameters remain same across all the transactions.
\begin{figure}
\centering
  \includegraphics[scale=0.4]{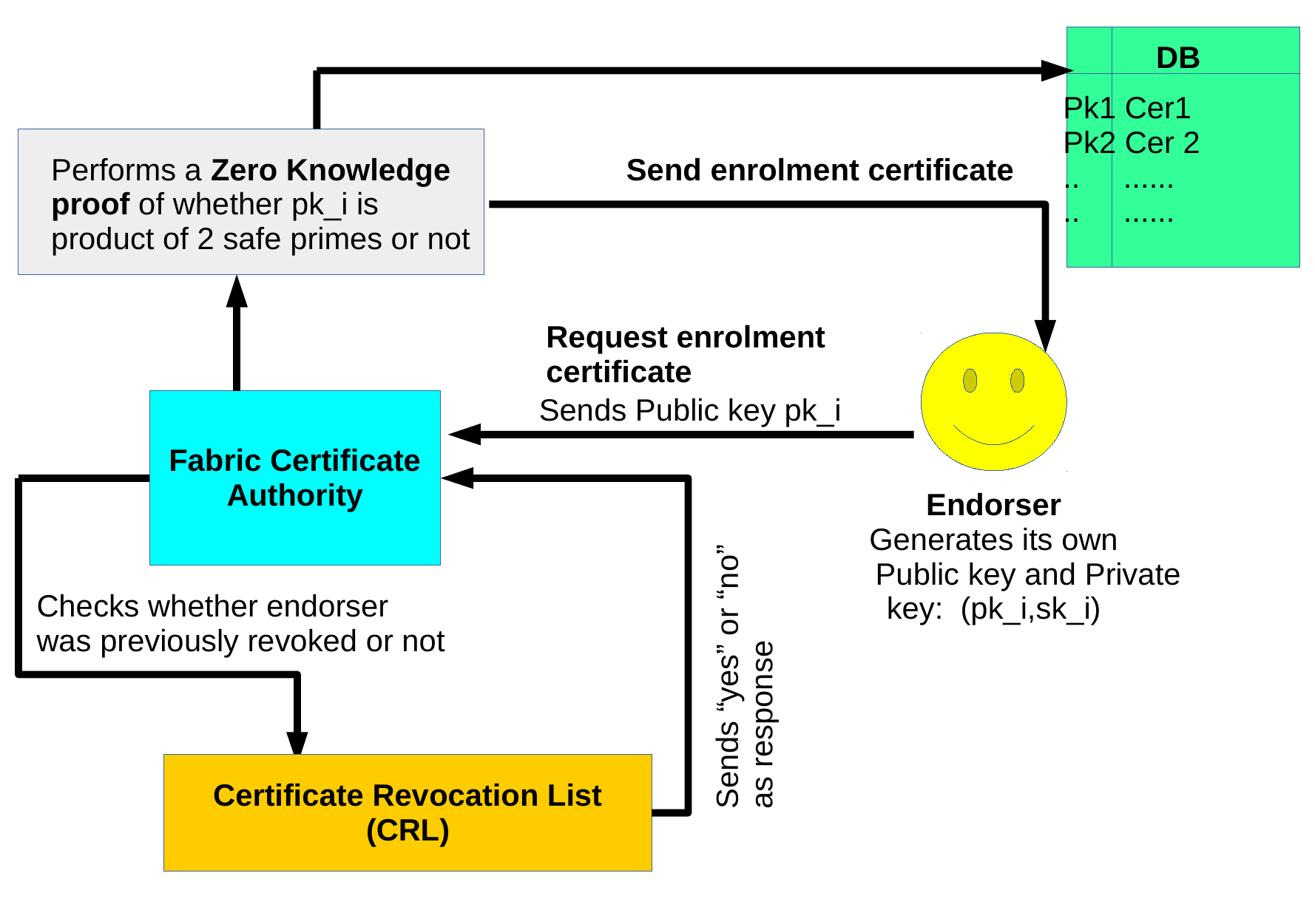}
  \caption{PKI}
  \label{fig:img2}

\end{figure}

\item[-] \textbf{\texttt{KeyGen}}. On input the system's parameters generated in \texttt{Init} phase, each endorser $E_i \in \mathcal{E},  1\leq i \leq n$ generates key pairs $(pk_i,sk_i)=(y_i,(p_i,q_i))$, $y_i=2p_iq_i+1,$ $p_i$ and $q_i$ being safe primes, by executing the probabilistic sampling algorithm $W$ of their accumulator\footnote{All endorsers run the sampling algorithm of the accumulator in parallel}. The range of $q_i \in S(2^{l/2},2^\mu)$. Upon obtaining the key pair, endorser $E_i$ submits its public key $y_i$ and \textit{verifiable credentials} to Fabric CA. The CA first checks whether such credentials matches with any of those present in \textit{Certificate Revocation List (CRL)}. If yes, then its \textit{enrolment certificate} was previously revoked and hence cannot be added as a network entity. Else, $E_i$ proves in zero-knowledge to CA the correctness of the value $y_i$\cite{camenisch1999proving}, \cite{camenisch1999separability}. If endorser is able to prove, then CA issues an \textit{enrolment certificate} to it. The identity of the endorser, public keys $y_j, 1 \leq j \leq n$ along with enrolment certificate gets added to the public database $\mathcal{DB}$ (any valid entity in the network has access to this database). The process has been illustrated in Fig \ref{fig:img2}.

\item[-] \textbf{\texttt{AccumulatePubKey}}. Fabric CA executes this algorithm for combining all the public keys in public database $\mathcal{DB}$. The accumulated value $v$ calculated by using data from $\mathcal{DB}$, is :
\begin{equation}
\begin{matrix}
v = f(u,\{y_j| 1 \leq j \leq n\}) \\
 =  f(f(\ldots f(u,y_1),y_2),y_3)\ldots,y_{n}) \\
 \ldots \ldots \\
= (\ldots(((u^{y_1} \mod N)^{y_2} \mod N)\ldots)^{y_{n}} \mod N)
 
  \end{matrix}
\end{equation}

This value is generated and used for long time unless the endorsement set $\mathcal{E}$ changes. Hence the computation can be said to be performed one time before instantiation of chaincode in all the peer nodes of the network.

\item[-] \textbf{\texttt{GeneratePubKeyWitness}}. Each member $e$ of set $\mathcal{E}$ computes witness $w_{e}\leftarrow f(u,\{y_i|1\leq i \leq n,i\neq e\}), \langle u \rangle = QR(N)$ for public key $y_{e}$, where accumulated value $v$ can be generated by computing $v \leftarrow f(w_e,y_e)$. When the endorser is willing to endorse or sign a transaction, it uses this value $w_e$ for construction of Signature based on Proof of Knowledge. As we have considered endorsement set to be fixed, even this value can be pre-computed.

\item[-] \textbf{\texttt{Sign}}. Endorser $E_{\pi} \in \mathcal{E}$ who wants to endorse a transaction is the Signer $\mathcal{S}$. It obtains the public key set $\mathcal{DB}=\{y_1,y_2,\ldots,y_n\}$, possessing a valid \textit{enrolment certificate} and has not been revoked (CA performs the check and informs if any endorser has been put in \textit{CRL}). 

A new linking criterion called \textit{Transaction-Oriented linkability} has been used in which one can tell if two signatures are linked if and only if they are signed by a common signer for a given transaction (similar to the concept of \textit{Event-oriented} linkability in \cite{tsang2004separable}). For this purpose we use a public parameter $g_{tid}$ instead of simply using $g \in QR(N)$. To construct $g_{tid}$, we consider $g \in QR(N)$ and a function $ \tilde{H}: \mathbb{N} \rightarrow \mathbb{G}$, $\mathbb{G} \subset \mathbb{Z}_{N/4}$ which generates $\tilde{tx}=\tilde{H}(\textrm{transaction-id})$, where \textit{transaction-id} is unique for each transaction, which is again the hash of the transaction payload \texttt{txPayload}. Thus, $g_{tid}=f(g,\tilde{tx})=g^{\tilde{tx}} \mod N$, where $f$ is the function defined as in Eq. \ref{eqn1}. 

For a given message $m \in \mathcal{M}$ (which is the \textit{transaction-response}) which has a transaction id \textit{transaction-id}, a private key $sk_\pi=(p_\pi,q_\pi)$ that corresponds to original public key, $y_\pi$, accumulated value $v$ and secret value $w_{\pi}$, signer $\mathcal{S}$ does the following (notations used as per \cite{au2006short}):

\end{itemize}
\begin{equation}
\label{eqsok}
SPK\Bigg\{ \Bigg(
\begin{tabular}{cc}
$w_\pi , y_\pi$ , \\
$p_\pi , q_\pi$ \\
\end{tabular}
\Bigg) : 
\begin{tabular}{cc}
$w_\pi^{y_\pi} = v \mod N  \wedge$ \\ $y_\pi = 2p_\pi q_\pi +1 \ \wedge \
y_\pi \in S(2^l,2^\mu)  $ \\ $\wedge \ q_\pi \in S(2^\frac{l}{2},2^\mu) \ \wedge$ \\
$\tilde{y}=\theta_d(p_\pi, q_\pi)$  \\
\end{tabular}
\Bigg\} (m)
\end{equation}
\begin{itemize}
\item[] where $\theta_d$ defined as $\theta_d(p_\pi, q_\pi)=g_{tid}^{p_\pi + q_\pi} \mod N$, is a one-way bijective mapping and $\tilde{y}$ is the tag generated corresponding to the signature.

\textit{``Signature based on Proof of Knowledge is basically a signature scheme in which a signer
can speak on behalf of any \textbf{NP} statement (as stated in \ref{eqsok}) to which he knows a witness/es without revealing all the irrelevant information \cite{chase2006signatures}.''} Here the witness values are $w_\pi,y_\pi,p_\pi$ and $q_\pi$. Any person who knows a satisfying assignment (that means posses the knowledge of witness) to the statements (in \ref{eqsok}) has signed the message.

\item[] A practical $\Sigma$-protocol for relation stated in Eq. \ref{eqsok} is constructed using the framework of discrete logarithm sets over group $QR(N)$.
The public parameters $g_{tid},h,y,t,s,\zeta \in QR(N)$ with unknown relative discrete logarithms alongwith the sequence of public values $T_1,T_2,T_3,T_4,T_5$ such that \\

$T_1=g_{tid}^r$ \ , $T_2=h^r \zeta^{x+r}$ \ , $T_3=s^r g_{tid}^{e_2}$ \ , $T_4=wy^r$, \ $T_5=t^r g_{tid}^{2e_1}$ \\

where $r \xleftarrow{R} [0,\lfloor N/4 \rfloor - 1]$ is used for the construction of proof.

The public values $T_1$ is for the free variable $r$, $T_2$ is for the free variable $x$, $T_3$ is for the free variable $e_2$, $T_4$ is for the free variable $w$ and $T_5$ is for the free variable $e_1$. Note that all the construction from $T_2$ to $T_5$ satisfy the property of triangularity with respect to first relation $T_1$. The construction of $T_2$ cannot be chosen of the form $h^r g_{tid}^x \mod N$ since $x$ belongs to the set $\mathcal{DB}$ whose size is negligible compared to the size $S(2^\frac{l}{2},2^\mu)$, exponential order of the security parameter $l$. If prover $P$ sends this value of $T_2$ to verifier $\mathcal{V}$, it can figure out, in polynomial time, the public key of the endorser/signer during verification phase.

The \textbf{NP} statements used for generating Signature based on Proof of Knowledge is given below:

\begin{equation}
\label{pok}
\begin{matrix}
T_1=g_{tid}^r , \ $(witness of $r$)$\\ 
T_2=h^r .\zeta ^{r}.\zeta^{x}  = h^r .\zeta ^{r+x}, \ $(witness of $x \in S(2^l,2^\mu)$)$ \\ 
(T_1)^{x} = g_{tid}^{a_1}, \ $(witness of $a_1$)$\\ 
(T1)^{e_2}=g_{tid}^{a_2},  \ $(witness of $a_2$)$\\ 
T_3=s^r g_{tid}^{e_2}, \ $(witness of $e_2 \in S(2^{l/2},2^\mu)$)$\\ 
(T_4)^{x}=vy^{a_1}, \ $(witness of $x$)$\\ 
(T_5)^{e_2}g_{tid}=t^{a_2}.g_{tid}^{x} \ $(witness of $e_2$ - a non-trivial factor of $x$)$\\ 
(T_3)^{2}T_5=s^{2r}.t^r.\tilde{y}^2  \ $(correctness of $\tilde{y}$)$\\ 
\end{matrix}
\end{equation}
 for the free variables $r,x,e_2,a_1,a_2$ such that $x \in \mathcal{S}(2^l,2^\mu), e_2 \in S(2^l,2^\mu), a_1=rx$ and $a_2=re_2$. The signer $\mathcal{S}$ gives a proof of knowledge for witness $w,y_\pi, p_\pi$ and $q_\pi$ by satisfying the above eight equations corresponding to the given accumulated value $v$. The variables $x,e_1 \ \textrm{and} \ e_2$ is assigned value $y_\pi$ (public key of $\mathcal{S}$), $p_\pi$ and $q_\pi$ respectively. 
It also proves that $(x - 1)/2$ can be factorized into two prime values, one of which belongs to $S(2^{l/2} , 2^\mu )$ and each of them are non trivial factors\cite{dodis2004anonymous}. 
 
Public Parameters : $g_{tid},h,t,s,y,\zeta,v \in QR(N)$, $T_1,T_2,T_3,T_4,T_5$,  $\mathbb{Z}_{N/4} \subset S(2^l,2^\mu)$.

\begin{itemize}
\item[1.] \texttt{Signer} $\mathcal{S}$ computes 
\begin{equation}
\label{sign}
\begin{aligned}
\alpha_1  &\xleftarrow{R} \mathbb{Z}_{N/4} & \alpha_2  &\xleftarrow{R} \mathbb{Z}_{N/4} \\
 \alpha_3  &\xleftarrow{R} \mathbb{Z}_{N/4},  & u_{1} &\leftarrow g_{tid}^{\alpha_1} \ mod \ N   \\
  u_{2} &\leftarrow \zeta^{\alpha_1+\alpha_2} \ mod \ N , &u_{3} &\leftarrow h^{\alpha_1} \ mod \ N, \\
  u_{4} &\leftarrow g_{tid}^{\alpha_1} \ mod \ N  &u_{5} &\leftarrow g_{tid}^{\alpha_3} \ mod \ N \\
   u_{6} &\leftarrow w^{\alpha_2}\ mod \ N, &u_{7} &\leftarrow g_{tid}^{2e_1.\alpha_3} \ mod \ N \\
   u_{8} &\leftarrow t^{\alpha_1} \ mod \ N , &u_{9} &\leftarrow g_{tid}^{\alpha_2} \ mod \ N\\
\end{aligned}
\end{equation}
\item[2.] $\mathcal{S}$ computes $c=H_1(m||u_1||u_2||u_3||u_4||u_5|| u_6||u_7||\\u_8||u_9), H_1: \mathcal{M} \times QR(N)^8 \rightarrow \mathcal{C}, \\ \mathcal{C} \subseteq QR(N)$ and uses it to compute 
\begin{equation}
\label{comm}
\begin{aligned}
\tilde{\alpha_1} &\leftarrow \alpha_1 + c.r  , &  \tilde{\alpha_2} &\leftarrow \alpha_2 + c.x \\
\tilde{\alpha_3} &\leftarrow r.\alpha_2 + r.c.x ,& \tilde{\alpha_4} &\leftarrow \alpha_3 + c.e_2 \\
 \tilde{\alpha_5} &\leftarrow r.\alpha_3 + r.c.e_2 & \\
\end{aligned}
\end{equation}
$\mathcal{S}$ sends the signature $\sigma'= (u_1,u_2,u_3,u_4,u_5,u_6,u_7,u_8,u_9,\tilde{\alpha_1}, \tilde{\alpha_2}, \tilde{\alpha_3}, \tilde{\alpha_4}, \tilde{\alpha_5},\tilde{y})$ where $\tilde{y}=g_{tid}^{p_\pi + q_\pi}$ to all the validators in set $\mathcal{V}.$  
\end{itemize}

\texttt{Signature size} : Values to communicated to the \textit{Verifier} as Signature based on Proof of Knowledge are $\sigma= (u_1,u_2,u_3,u_4,u_5,u_6,u_7,u_8,u_9,\tilde{\alpha_1},\tilde{\alpha_2}, \tilde{\alpha_3}, \tilde{\alpha_4},\tilde{\alpha_5},\tilde{y})$, $ \\ u_1,u_2,\ldots,u_9, \tilde{y}, \tilde{\alpha_1},\tilde{\alpha_2} \ \textrm{and} \ \tilde{\alpha_4}$, each approximately being $\lambda$ bits in size and $\tilde{\alpha_3},  \tilde{\alpha_5}$ each approximately being $2\lambda$ bits in size. Hence the signature generated is of constant size, being $\mathcal{O}(\lambda)$ where $\lambda$ is the security parameter. 

\item[-] \textbf{\texttt{Verify}}. To verify the signature \\ $\sigma'= (u_1,u_2,u_3,u_4,u_5,u_6,u_7,u_8,u_9,\tilde{\alpha_1},\tilde{\alpha_2},\tilde{\alpha_3}, \tilde{\alpha_4},\tilde{\alpha_5},\tilde{y})$ on message $m \in \mathcal{M}$, all the validator nodes in $\mathcal{V}$ computes $c=H_1(m||u_1||u_2||u_3|| u_4||u_5||u_6||u_7||u_8||u_9), \\ H_1: \mathcal{M} \times QR(N)^8 \rightarrow \mathcal{C}, \mathcal{C} \subseteq QR(N)$ and checks if all the statements in Eq. \ref{ver} is valid or not. For \textit{1-out-of-n} endorsement policy, if all check passes, then the verifier outputs \texttt{accept}; otherwise it outputs \texttt{reject} and aborts. For \textit{t-out-of-n}, $t>1$, we need to perform the test for signature linkability (\texttt{Link}) as well for final acceptance.

\item[-] \textbf{\textbf{Link}}. In \cite{au2006short}, the tag generated is $\theta_d=((e 1 , e 2 )) = g^{e_1 +e_2}$. $\theta_d$ being PK-bijective, it prevented double signing on the same message. However as the endorsement set remains fixed,  tag constructed must be function of the secret key as well as the transaction payload. For (\textit{Transaction-Oriented linkability}), tag construction is modified by introducing transaction-id, a unique value associated with each transaction payload. Since $g_{tid} \in QR(N)$, modified $\theta_d=g_{tid}^{e_1+e_2}$ remains PK-bijective. 
Given two valid signatures $\sigma_1'$ and $\sigma_2'$ for a given transaction, validator node checks if $\tilde{y_1} = \tilde{y_2}$. If yes, output \textbf{linked}. Otherwise, output \textbf{unlinked}. 
\end{itemize}
\subsection{Extending to threshold endorsement policy}
\label{thres}
Given a threshold value $t$, if a validator node receives at least \textit{t} out of $n$ \textit{transaction-response} with a valid, pairwise unlinked signatures (after $\big( ^t _2 \big)$ tests of linkability) whose responses (read set and write set) are the same, then the endorsement policy is said to be satisfied. If each of at least $\frac{|\mathcal{V}|}{2}$ validator nodes in $\mathcal{V}$ reach a consensus on receipt of at least \textit{t} signatures for the given transaction, then one of the honest validator node ``broadcast'' the \textit{transaction-response} within a \textit{transaction message} to the \textit{ordering service} so that the transactions can be ordered chronologically by the channel. 

\begin{equation}
\label{ver}
\begin{matrix}
g_{tid}^{\tilde{\alpha_1}} \stackrel{?}{=} u_{1}.T_1^c ,\\
g_{tid}^{\tilde{\alpha_1}} \stackrel{?}{=} g_{tid}^{\alpha_1}.{g_{tid}^r}^c \mod N \ (\because Eq. \ \ref{pok},\ref{sign},\ref{comm}), \\
g_{tid}^{\tilde{\alpha_1}} \stackrel{?}{=} g_{tid}^{\alpha_1 + r.c} \mod N\\
\\
\zeta^{\tilde{\alpha_2}+\tilde{\alpha_1}} h^{\tilde{\alpha_1}} \stackrel{?}{=} u_{2}.u_3.T_2^c ,\\
\zeta^{\tilde{\alpha_1}+\tilde{\alpha_2}} h^{\tilde{\alpha_1}} \stackrel{?}{=} \zeta^{\alpha_1+\alpha_2}.h^{\alpha_1}.{(h^r.\zeta^{x+r})}^c \mod N, \ (\because Eq. \ \ref{pok},\ref{sign},\ref{comm}), \\
\zeta^{\tilde{\alpha_1}+\tilde{\alpha_2}} h^{\tilde{\alpha_1}} \stackrel{?}{=} \zeta^{\alpha_2 + \alpha_1 + c.(x + r)}.h^{\alpha_1 + r.c} \mod N \\
\\
g_{tid}^{\tilde{\alpha_3}} \stackrel{?}{=} T_1^{\tilde{\alpha_2}},\\
g_{tid}^{\tilde{\alpha_3}} \stackrel{?}{=} {(g_{tid}^r)}^{\alpha_2 + c.x} \mod N, \ (\because Eq. \ \ref{pok},\ref{comm}), \\
g_{tid}^{\tilde{\alpha_3}} \stackrel{?}{=} g_{tid}^{r.\alpha_2 + r.c.x} \mod N.\\
\\
g_{tid}^{\tilde{\alpha_5}} \stackrel{?}{=} T_1^{\tilde{\alpha_4}},\\
g_{tid}^{\tilde{\alpha_5}} \stackrel{?}{=} (g_{tid}^r)^{\alpha_3 + c.e_2} \mod N \ (\because of Eq. \ref{pok},\ref{comm}),\\
g_{tid}^{\tilde{\alpha_5}} \stackrel{?}{=} g_{tid}^{r.\alpha_3 + r.c.e_2} \mod N .\\
\\
g_{tid}^{\tilde{\alpha_4}}.s^{\tilde{\alpha_1}} \stackrel{?}{=} T_3^c.u_4.u_5 ,\\
g_{tid}^{\tilde{\alpha_4}}.s^{\tilde{\alpha_1}} \stackrel{?}{=} (s^r g_{tid}^{e_2})^c.s^{\alpha_1}.g_{tid}^{\alpha_3} \mod N \ (\because Eq. \ \ref{pok},\ref{sign},\ref{comm}),\\
g_{tid}^{\tilde{\alpha_4}}.s^{\tilde{\alpha_1}} \stackrel{?}{=} g_{tid}^{\alpha_3 + c.e_2}.s^{\alpha_1 + c.r}  \mod N\\
\\
u_6.v^c.y^{\tilde{\alpha_3}} \stackrel{?}{=} T_4^{\tilde{\alpha_2}}, \\
w^{\alpha_2}.(w^x)^c.y^{\tilde{\alpha_3}} \stackrel{?}{=} (w.y^r)^{\tilde{\alpha_2}} \mod N \ (\because Eq. \ \ref{pok},\ref{sign},\ref{comm}),\\
w^{\alpha_2+c.x}.y^{\tilde{\alpha_3}} \stackrel{?}{=} w^{\tilde{\alpha_2}}.y^{r.\tilde{\alpha_2}} \mod N\\
\\
 
t^{\tilde{\alpha_5}}.g_{tid}^{\tilde{\alpha_2}}.u_7 \stackrel{?}{=} T_5^{\tilde{\alpha_4}}.u_9.g_{tid}^c \\
t^{\tilde{\alpha_5}}.g_{tid}^{\tilde{\alpha_2}}.g_{tid}^{2.e_1.\alpha_3} \stackrel{?}{=} (t^r.g_{tid}^{2.e_1})^{\tilde{\alpha_4}}.g_{tid}^{\alpha_2}.g_{tid}^c \mod N \ (\because Eq. \ \ref{pok},\ref{sign},\ref{comm}),\\
t^{\tilde{\alpha_5}}.g_{tid}^{\tilde{\alpha_2} + 2.e_1.\alpha_3 } \stackrel{?}{=} t^{r.\tilde{\alpha_4}}g_{tid}^{2.e_1.(\alpha_3+c.e_2) + \alpha_2 + c} \mod N\\
t^{\tilde{\alpha_5}}.g_{tid}^{\tilde{\alpha_2} + 2.e_1.\alpha_3 } \stackrel{?}{=} t^{r.\tilde{\alpha_4}}g_{tid}^{2.e_1.\alpha_3+ 2.c.e_1.e_2 + \alpha_2 + c} \mod N\\
t^{\tilde{\alpha_5}}.g_{tid}^{\tilde{\alpha_2} + 2.e_1.\alpha_3 } \stackrel{?}{=} t^{r.\tilde{\alpha_4}}g_{tid}^{2.e_1.\alpha_3+ c.(x-1) + \alpha_2 + c} \mod N\\
\\

\tilde{y}^{2c}.s^{2\tilde{\alpha_1}}.t^{\tilde{\alpha_1}} \stackrel{?}{=} (T_3^2. T_5)^c. u_4^2. u_8  \\
 \tilde{y}^{2c}.s^{2\tilde{\alpha_1}}.t^{\tilde{\alpha_1}} \stackrel{?}{=} ((s^r.g_{tid}^{e_2})^2. t^r.g_{tid}^{2.e_1})^c. (s^{\alpha_1})^2. t^{\alpha_1} \mod N  (\because Eq. \ \ref{pok},\ref{sign},\ref{comm}),\\
 \tilde{y}^{2c}.s^{2\tilde{\alpha_1}}.t^{\tilde{\alpha_1}} \stackrel{?}{=} s^{2.r.c}.g_{tid}^{2.c.e_2}. t^{r.c}.g_{tid}^{2.e_1.c}.s^{2.\alpha_1}. t^{\alpha_1} \mod N\\ 
 \tilde{y}^{2c}.s^{2\tilde{\alpha_1}}.t^{\tilde{\alpha_1}} \stackrel{?}{=} g_{tid}^{2.c.e_1 + 2.c.e_2}.s^{2.r.c + 2.\alpha_1}. t^{r.c + \alpha_1} \mod N\\ 

\end{matrix}
\end{equation}

\section{Security Model}
\label{Chap:security}
\subsection*{Assumptions made}
Some assumptions made regarding the entities in Hyperledger Fabric are - all communication channels are secure, Fabric CA is honest, members of endorsement set is fixed , all the peer nodes have their local copy of database consistent with the world state, signature generation algorithm follows a \textit{Random Oracle Model}. In order to satisfy the threshold endorsement policy when at least $t$ signers are willing to endorse, we assume that at least half of the members in the verifier set and more than half (at least $n/2+1$) endorsers in an endorsement set is honest.
The security model defined here is similar to the one defined in \cite{tsang2005short},\cite{au2006short}.
\subsection*{Syntax}
A \textit{Linkable Ring Signature} scheme is a tuple (\textbf{Init, KeyGen, AccumulatePubKey, GeneratePubKeyWitness, Sign, Verify, Link}) of seven polynomial time algorithms. Instead of a single entity generating keys for all the participants in the permissioned blockchain, we define \textbf{KeyGen} as an algorithm executed by each individual user for the generation of the public and private key pair. Syntax is as follows :
\begin{itemize}
\item[-] \textbf{param} $\leftarrow$ \textbf{Init}($1^\lambda$), the poly-time \textit{initialization} algorithm which, on input a security parameter $\lambda \in \mathbb{N}$, outputs the system parameters containing, among other things, $1^\lambda$. All other algorithms implicitly use $\lambda$ as one of their inputs.
\item[-] \textbf{($sk_i,pk_i$)} $\leftarrow$ \textbf{KeyGen()}, the PPT (\textit{probabilistic polynomial time}) \textit{key generation} algorithm which outputs a secret/public key pair ($sk_i,pk_i$). $\mathcal{SK}$ and $\mathcal{PK}$ denote the domains of possible secret keys and public keys respectively. All the generated $pk_i, 1\leq i \leq n$ for $n$ participants is made publicly available along with system parameters.  
\item[-] \textbf{($v$)} $\leftarrow$ \textbf{AccumulatePubKey()}, the deterministic poly-time algorithm which, on input a set $\mathcal{Y}$ of $n$ public keys in $\mathcal{PK}$, where $n \in \mathbb{N}$ is of size polynomial in $\lambda$, produces the value $v$.
\item[-] \textbf{($w$)} $\leftarrow$ \textbf{GeneratePubKeyWitness()}, the deterministic poly-time algorithm which, on input a set $\mathcal{Y'}=\mathcal{Y}\setminus \{pk_e\}$ i.e. all public keys except that of the entity $e$ who executes it ($n \in \mathbb{N}$ is of size polynomial in $\lambda$), produces the value $w_e$ such that $f(w_e,pk_e)=w_e^{pk_e}=v$.
\item[-]  For a \textit{Signatures based on Proofs of Knowledge}, the $\Sigma$-protocol between signer and verifier for the \textbf{NP}-relation stated in Eq.\ref{eqsok} has been converted into a signature scheme. It comprises the \textbf{(Sign,Verify)} algorithm pair, executed on the signer and verifier side respectively. Execution of this protocol is time independent from the number of public keys that gets aggregated in \textbf{AccumulatePubKey} or \textbf{GeneratePubKeyWitness}. 
\begin{itemize}
\item[-] $\sigma \leftarrow$ \textbf{Sign($\mathcal{Y},M,x$)}, the PPT \textit{signing} algorithm which, on input a set $\mathcal{Y}$ of $n$ public keys in $\mathcal{PK}$, where $n \in \mathbb{N}$ is of size polynomial in $\lambda$, a message $M \in \{0,1\}^*$, and a private key $x \in \mathcal{SK}$ whose corresponding public key is contained in $\mathcal{Y}$, produces a signature $\sigma$. We denote by $\Sigma$ the domain of possible signatures.
\item[-] 1/0 $\leftarrow$ \textbf{Verify($\mathcal{Y},M,\sigma$)}, the poly-time \textit{verification} algorithm which, on input a set $\mathcal{Y}$ of $n$ public keys in $\mathcal{PK}$, where $n \in \mathbb{N}$ is of size polynomial in $\lambda$, a message $M \in \{0,1\}^*$ and a signature $\sigma \in \Sigma$, returns 1 or 0 meaning \textbf{accept} or \textbf{reject} respectively. If the algorithm returns \textbf{accept}, the message-signature pair $(M,\sigma)$ is said to be \textit{valid}.
The signature scheme must satisfy \textit{Verification Correctness}, i.e. signatures signed by honest signer as per the specification must be accepted by an honest verifier with overwhelming probability.
\end{itemize}
\item[-] 1/0 $\leftarrow$ \textbf{Link($\sigma_0,\sigma_1$)}, the poly-time linking algorithm which, on input two valid signatures, checks their corresponding tag and outputs 1 (if tags are same - signatures are linked) or 0 (if tags are different - unlinked signature) meaning \textbf{linked} or \textbf{unlinked} respectively. The signature scheme must satisfy \textit{Linking Correctness}, i.e. any two signatures signed by a common honest signer on the same message are \textbf{linked} with overwhelming probability. On the other hand, any two signatures signed by two different honest signer must be \textbf{unlinked} with overwhelming probability.
\end{itemize}
\subsection*{Security Notions}
Before defining the security notions, let us define the adversarial model and the possible attacks :
\subsection{Adversarial Model}
\begin{itemize}
\item Any corrupt endorser may launch insider attack (threat or use of influence) on the rest of the endorsers, acquiring their private keys.
\item Members (excluding the signer) belonging to the endorsement set may collude and reveal their secret keys on receipt of signature.
\item Validator may hold back the packets without verifying.
\item Validator can act maliciously by randomly mark a transaction as valid/invalid without actually verifying.
\end{itemize}
Any adversary is assumed to have the following oracle access:
\begin{itemize}
\item The \textit{Corruption Oracle,} which outputs the corresponding secret key given a public key as input. 
\item The \textit{Signing Oracle,} which returns a valid signature, on input a designated signer $s$, message $M$ and subring $R$ (comprising subset of public keys). The signature is computationally indistinguishable from one produced by \texttt{Sign($\mathcal{Y},M,x$)} using the real secret key $x$ of signer $s$ on message $M$, $\mathcal{Y}$ being the set of all public keys. 
\end{itemize} 


Note, that if endorsement logic gets corrupted by adversary then it needs a mechanism of formal verification to check whether desired output is achieved or not. This is beyond our scope of work. Based on the last two points, we discus the correctness and soundness property of the $\Sigma$-protocol as well as security of signature scheme in the permissioned blockchain framework.

\subsection{Correctness}
For correctness, any execution of the $\Sigma$-protocol for the NP-relation given in Eq.\ref{eqsok} will terminate with the verifier outputting 1, with overwhelming
probability, if and only if a prover or an endorser possess the correct witness values  $(y_\pi, w_\pi, p_\pi,q_\pi)$ for the corresponding accumulated public value $v$.
\subsection{Soundness}
The Honest-Verifier Zero-Knowledge property of the $\Sigma$-protocol for \textbf{NP}-relations stated in Eq.\ref{eqsok} guarantees that the transcript generated out of the interaction between signer and verifier does not leak any information to the adversary $\mathcal{A}$ that has no knowledge of the secret. The soundness property is formalized in terms of the game played between Fabric CA and adversary $\mathcal{A}$, assuming all endorsers, participating in ring formation, are honest.
\begin{itemize}
\item Fabric CA runs the \texttt{Init} algorithm for security parameter $\lambda$ and generates system parameters. All endorsers executes \texttt{KeyGen} algorithm to generate the public key and private key pair and stored in public database $\mathcal{DB}$.
\item $\mathcal{A}$ receives system parameters from Fabric CA and gets the transcript of prior runs of the protocol between an honest signer and verifier. Given $\mathcal{A}$ has access to the corruption oracle, it can query for the secret key of some but not all endorsers, who has put their public keys in database $\mathcal{DB}$. 
\item $\mathcal{A}$ now select a set of endorsers $E'$ for which it has not queried their secret keys. It generates a value $v'$ by accumulation of public keys of $E'$.
\item $\mathcal{A}$ starts executing the $\Sigma$-protocol in the role of the signer and the probability of winning the game is negligible. Following the correctness property, an honest verifier with output 1(accept) with overwhelming probability if and only if the $\mathcal{A}$ can produce the correct secret value $(y_\pi,w_\pi,p_\pi,q_\pi)$ corresponding to accumulated value $v'$.  
\end{itemize}

Note that if $\mathcal{A}$ is not given access to a correct tuple $(y_\pi,w_\pi,p_\pi,q_\pi)$ but still it wins the game, then it must have generated it by himself/herself. This contradicts the one-wayness of accumulator's domain.

\subsection{Security Analysis of Signature Scheme}
Since the architecture of \textit{Fabric} is modular, the proposed signature scheme can be plugged-in as a feature. Given that Fabric is secure (Security model discussed in \cite{Hyperledger},\cite{androulaki2018hyperledger}), we need to argue on the security of the proposed anonymous endorsement system based on the security of \textit{FCsLRS} scheme.
\begin{theorem}
 If FCsLRS scheme is unforgeable, linkable anonymous, linkable and non-slanderable, then the scheme is secure and hence the proposed Anonymous Endorsement System also remains secure in the random oracle model.
\end{theorem}

We have defined the security notions in details :
\subsubsection*{Unforgeability} 
The following construction of \textit{constant-sized linkable ring signature} is unforgeable against ``chosen endorser" attacks (means a subset of set $E$ is selected and only the public keys of those endorsers are taken into consideration). The adversary is further allowed to corrupt endorsers and acquire their private keys using \textit{corruption oracle}.
\begin{definition}
\label{def2}
\textbf{(Unforgeability)}.\cite{au2006short} \textit{Any corrupt endorser acting as adversary $\mathcal{A}$ is given access to the public keys of all the members belonging to endorsement set $E$ as well as the \textit{signing oracle} and \textit{corruption oracle}. $\mathcal{A}$ is allowed to query the \textit{signing oracle} for signature on message of its choice for a given subset of endorsers and use corruption oracle to get secret key for a set of corrupt endorsers denoted by $C : C \subset E$. A linkable ring signature scheme is unforgeable if for any PPT adversary $\mathcal{A}$ and for any polynomial n(.), the probability that $\mathcal{A}$ succeeds in forging signature on a message which it has not queried before, for a set of endorsers including at least more than one honest member, is negligibly close to $1/2$.}
\end{definition}
%
\subsubsection*{Linkable-Anonymity}

If all the endorsers except one honest signer gets corrupted and reveals their secret key in order to frame the signer (\cite{au2006short}), the scheme does not ensure anonymity anymore. But in FCsLRS, for \textit{t-out-of-n} threshold endorsement policy, since more than half of the endorsers in $\mathcal{E}$ is assumed to be honest, possibility of such attacks is negligible. At least $n/2+1$ members in $\mathcal{E}$ will not reveal their secret keys . %

\begin{definition} 
\label{la}
(\textbf{Linkable-anonymity}).\cite{au2006short} \textit{Assuming that any corrupt endorser acting as adversary $\mathcal{A}$ has the same advantage of obtaining signature on message of its choice and querying for secret keys as stated in Definition \ref{def2}. $\mathcal{A}$ now selects two public keys $PK_{i_0},PK_{i_1}$ of its choice (which has not been used for querying the signing oracle or corruption oracle), a message $M$ and subset of $E$ denoted by $E_{sub}$. The challenger now selects any public key at random out of $PK_{i_0}, PK_{i_1}$ and generates a signature of $M$ over set $E_{sub}$. After this step, $\mathcal{A}$ is allowed to query the signing oracle and corruption oracle  for any public key except $PK_{i_0}$ and $PK_{i_1}$. A linkable ring signature is $\textrm{linkably anonymous}$, if for any PPT adversary $\mathcal{A}$ and for any polynomial n(.), the probability that $\mathcal{A}$ succeeds in guessing the correct public key (either $PK_{i_0}$ or $PK_{i_1}$), for which the signature was generated by challenger, is negligibly close to $1/2$.}
 
\textit{Adversarially-chosen keys defines the power of $\mathcal{A}$ which allows it to select public keys outside the set $E$ for constructing set $E_{sub}$ (but $PK_{i_0},PK_{i_1} \in E$) and allowing use of such externally chosen public keys for querying the signing oracle. A linkable ring signature is linkably anonymous w.r.t adversarially-chosen keys, if for any PPT adversary $\mathcal{A}$ and for any polynomial n(.), the probability that $\mathcal{A}$ succeeds in guessing the correct public key (either $PK_{i_0}$ or $PK_{i_1}$), for which the signature was generated by challenger, is negligibly close to $1/2$.}
\end{definition}
\subsubsection*{Linkability}  

Under the SRSA assumption of \textit{Accumulators with one-way domain}, it is hard to generate the secret keys $(e_1,e_2)$ for a given public key value $x$ of an endorser. Thus the probability of producing a valid signature for a given transaction(message) and secret key pair is negligible. Security is ensured even in presence of adversarially-chosen keys. 

\begin{definition}(\textbf{Linkability}).\cite{au2006short}
\label{linkable}
\textit{Assuming that any corrupt endorser acting as adversary $\mathcal{A}$ has the same advantage of obtaining signature on message of its choice and querying for secret keys as stated in Definition \ref{def2}. A linkable ring signature is linkable if for any PPT adversary $\mathcal{A}$ and for any polynomial n(.), the probability that $\mathcal{A}$ succeeds in generating two different signature $\sigma_1,\sigma_2$ for the same message over same set of endorsers defined by $E$ without getting linked (by returning $Link(\sigma_1,\sigma_2)=0$), is negligibly close to $1/2$.}

The signature scheme is also linkable w.r.t to adversarially-chosen keys.
\end{definition}

%
\subsubsection*{Non-slanderability} 
It ensures that any corrupt endorser cannot produce a linkable signature on behalf of or frame an honest endorser(\cite{wei2005tracing}). For our construction of FCsLRS, we consider a tag generation which provides \textit{Transaction-Oriented} linkability. Since each tag generated is dependent on the transaction id and secret key pair and transaction id being unique for each transaction\footnote{transaction-id is the hash of the transaction payload} and assuming the function $\tilde{H}$ is a random oracle , it is hard to produce two linked signatures for same transaction. Security is ensured even in presence of adversarially-chosen keys. 

\begin{definition}
(\textbf{Non-slanderability}).\cite{au2006short} \textit{Assuming that any corrupt endorser acting as adversary $\mathcal{A}$ has the same advantage of obtaining signature on message of its choice and querying for secret keys as stated in Definition \ref{def2}. Given that an honest member of $E$ with public key $PK$ can generate a valid signature $\sigma$ on a message $M$ for subset of $E$, a linkable ring signature is non-slanderable if for any PPT adversary $\mathcal{A}$ and for any polynomial n(.), the probability that $\mathcal{A}$ succeeds in the generating a valid signature $\sigma^*$ corresponding to the public key $PK$ (provided this public key was not used while querying corruption oracle or signing oracle) such that $Link(\sigma^*,\sigma)=0$, is negligibly close to $1/2$.}

\end{definition}
The signature scheme is also non-slanderable w.r.t to adversarially-chosen keys.

Now we define the theorems that justifies the security of FCsLRS and provide the security proofs as well :
\begin{theorem}
\label{th1}
\textit{``If the DDH in QR(N) problem, the LD-RSA problem, the Strong-RSA
problem are hard and the function $\tilde{H}$ is random oracle, our construction is unforgeable. \cite{au2006short}''}
\end{theorem}
\textit{Proof}. If the signature scheme is \textit{Non-Slanderable} and \textit{Linkable} then it is \textit{Unforgeable}. That is, if a corrupt endorser or malicious peer node can forge a signature, then he can even frame an honest endorser and sign on his behalf or collude with other corrupted endorsers to break the linkability of the signature. Therefore, theorems \ref{th3} and \ref{th4} together imply that the scheme is unforgeable.

\begin{theorem}
\label{th2}
\textit{``Under the assumption that endorsement set has at least two honest members, if the DDH in QR(N) problem, the LD-RSA problem, the Strong-RSA
problem are hard and the function $\tilde{H}$ is random oracle ,
then our construction is linkably-anonymous w.r.t. adversarially-chosen keys.\cite{au2006short}''}
\end{theorem}
\textit{Proof Sketch}. Here we proof the contrapositive of the theorem, i.e. if we can construct a simulator $S$ from a corrupt endorser $\mathcal{A}$ which succeeds in guessing the correct public key (either $PK_{i_0}$ or $PK_{i_1}$) corresponding to the signature generated by challenger (as defined in \textit{Definition} \ref{la}), then it can also solve the LD-RSA problem under the DDH assumption.
\par Given the values $n_0=p_0.q_0,n_1=p_1.q_1$, a bit $b \in_R \{0,1\}$ is selected randomly and the value $Tag=g_{tid}^{p_b+q_b}$ is generated by challenger. The values $(n_0,n_1,Tag)$ is transmitted to $S$. Using the system parameters, it randomly generate a set of key pairs $J = \{(PK_i , SK_i )\}$ $1\leq i \leq |E|$ in order to simulate the actual situation. Now $S$ selects a bit $b' = 0$ or 1 randomly and sets $PK^*$ to $2n_{b'} + 1$. The set of public keys $J^* = J \cup \{ PK^* \}$ is then given to $\mathcal{A}$.


\par The adversary $A$ may query the signing oracle for any of the public keys present in $J^*$. If the public key belongs to set $J$, then simulator can straight away generate the signature, given that it posseses the secret key. If the request comes from a public key not present in $J$ (but not $PK^*$), $S$ first asks the adversary to submit a proof showing that it has correctly generated the key pair. Here, $S$ can extract the secret key during the proof of validity of public key. Now the signature generation can proceed as was stated before. If the request comes for public key $PK^*$, then $S$ sets tag $\tilde{y} = Tag$ and computes the signature of knowledge. The simulated signature of
knowledge (constructed as in Eq. \ref{eqsok}.) is indistinguishable from the actual one under the DDH assumption in $QR(N)$, provided $Tag$ is correctly formed. This is possible if and only if the bit $b$ selected by challenger is same as bit $b'$. Since the advantage of breaking linkable anonymity is non-negligibly more than $1/2$, advantage of $S$ in generating a valid signature of knowledge will also be non-negligibly more than $1/2$, thereby solving the LD-RSA problem.
\begin{theorem}
\label{th3}
\textit{``If the DDH in QR(N) problem, the LD-RSA problem, the Strong-RSA
problem are hard and the function $\tilde{H}$ is random oracle, then our construction is linkable w.r.t. adversarially-chosen public keys of endorsers.\cite{au2006short}''}
\end{theorem}
\textit{Proof Sketch}. In order to break the linkability property, a corrupt endorser $\mathcal{A}$ has to convince a verifier to accept a invalid tag $\tilde{y}$ with non-negligible probability. This is possible if $\mathcal{A}$ is successful in forging a signature for a given message over a given set of endorsers or it can generate an incorrect proof of construction for the invalid tag \cite{dodis2004anonymous}.  
Either of the case is not possible, since forging a signature is hard under the Strong-RSA assumption and generating an incorrect proof is hard under LD-RSA assumption. 
\begin{theorem}
\label{th4}
\textit{``If the DDH in QR(N) problem, the LD-RSA problem, the Strong-RSA
problem are hard and the function $\tilde{H}$ is random oracle, then our construction is non-slanderable w.r.t. adversarially-chosen public keys of endorsers.\cite{au2006short}''}
\end{theorem}
\textit{Proof Sketch}. If a corrupt endorser $\mathcal{A}$ is able to output a signature which slanders an honest endorser with public key $PK^*$ and the soundness property of \textit{Signature based on Proof of Knowledge} (Eq. \ref{eqsok}) holds, then there exists a \textit{knowledge extractor (def. \ref{def:ke})} which can extract the secret key $(p',q')$ corresponding to the public key $PK^*$  such that $PK^* = 2p'q'+ 1$. Hence simulator S solves the LD-RSA problem. 

\section[Performance Analysis of FCsLRS]{Performance Analysis of FCsLRS \footnote{Over here we analyze the performance for generation and verification of one signature}}
\label{Chap:performance}
\noindent
The (\textbf{Sign,Verify}) algorithm pair just involves the execution of the $\Sigma$-protocol which is time independent from the number of public keys that were aggregated when constructing the accumulated value $v$ and generation of witness value $w$. At the end of each protocol run, verifier $\mathcal{V}$ outputs a 0/1.
\subsection{Theoretical Analysis}
 \begin{table}[H]
    \caption{Asymptotic complexity analysis of FCsLRS }  
  \centering
  \begin{tabular}{|c |c | c|}
   \hline
  Algorithm &Operations performed  &Asymptotic complexity  \\
 \hline
 Tag generation &$2E$ & $\mathcal{O}(\lambda)$\\ 
 Signature &$11E + 5M$ &$\mathcal{O}(\lambda)$\\
 Verification &$6E + 10M$ &$\mathcal{O}(\lambda)$\\
\hline
\end{tabular}
\label{asym}
\end{table}
\textcolor{black}{Table \ref{asym}. provides an asymptotic analysis of the signature scheme where we consider the following mathematical operations based on the Eq.\ref{sign}, \ref{comm} and \ref{ver} -} \textit{E} is an exponentiation (single base of the form  $g^a$ ) operation, \textit{M} is multibase exponentiation (of the form  $g^a.h^b$) operation and $\lambda$ is the security parameter.
\subsection{Experimental Analysis}
\label{exp-analysis}
The performance of the signature scheme was measured on \texttt{Intel Core i5-4200U CPU, quad core processor, frequency 1.60 GHz}, OS : \textit{Ubuntu-16.04 LTS} (64 bit). The programming language used is \texttt{Go 1.10} \cite{golang}, packages used is \textit{crypto,   golang.org/x/ crypto/sha3, rand} and \textit{math}. The code for \textit{Cyclic Group Generator} is based on the one given under \texttt{Project-iris}\footnote{https://github.com/project-iris/iris/blob/v0.3.2/crypto/cyclic/cyclic.go}.  
For the analysis of the signature generation and verification time, the RSA modulus size, also the security parameter $\lambda$, has been varied as 1024 bits, 2048 bits and 3072 bits. The endorsement set size (number of participants, $n$) was varied as $4,8,\ldots,256$, in ascending powers of 2 and message length was varied as 2KB, 4KB and 8KB respectively. The functions $H_1$ and $\tilde{H}$ used is \textit{SHA-3} producing a message digest of size 128 bits. The range of both the hash functions must be a subset of $QR(N), N$ is the $RSA$-modulus. Also as per Eq (4), any element in the group $QR(N)$ when raised to the power of a number in the group $\mathbb{Z}_{N/4}$, again returns an element which belongs to $QR(N)$. Since  $N$ is varied between 1024 bits to 3072 bits and least value of $N/4$ is 256 bits, so a message digest of 128 bit is definitely falling in the range $\mathbb{Z}_{N/4}$ for any case. We had to do this approximation since selection of an element from $QR(N)$ when $\phi(N)$ is unknown is hard. 

\begin{figure}[H]
\centering
  \includegraphics[scale=0.6]{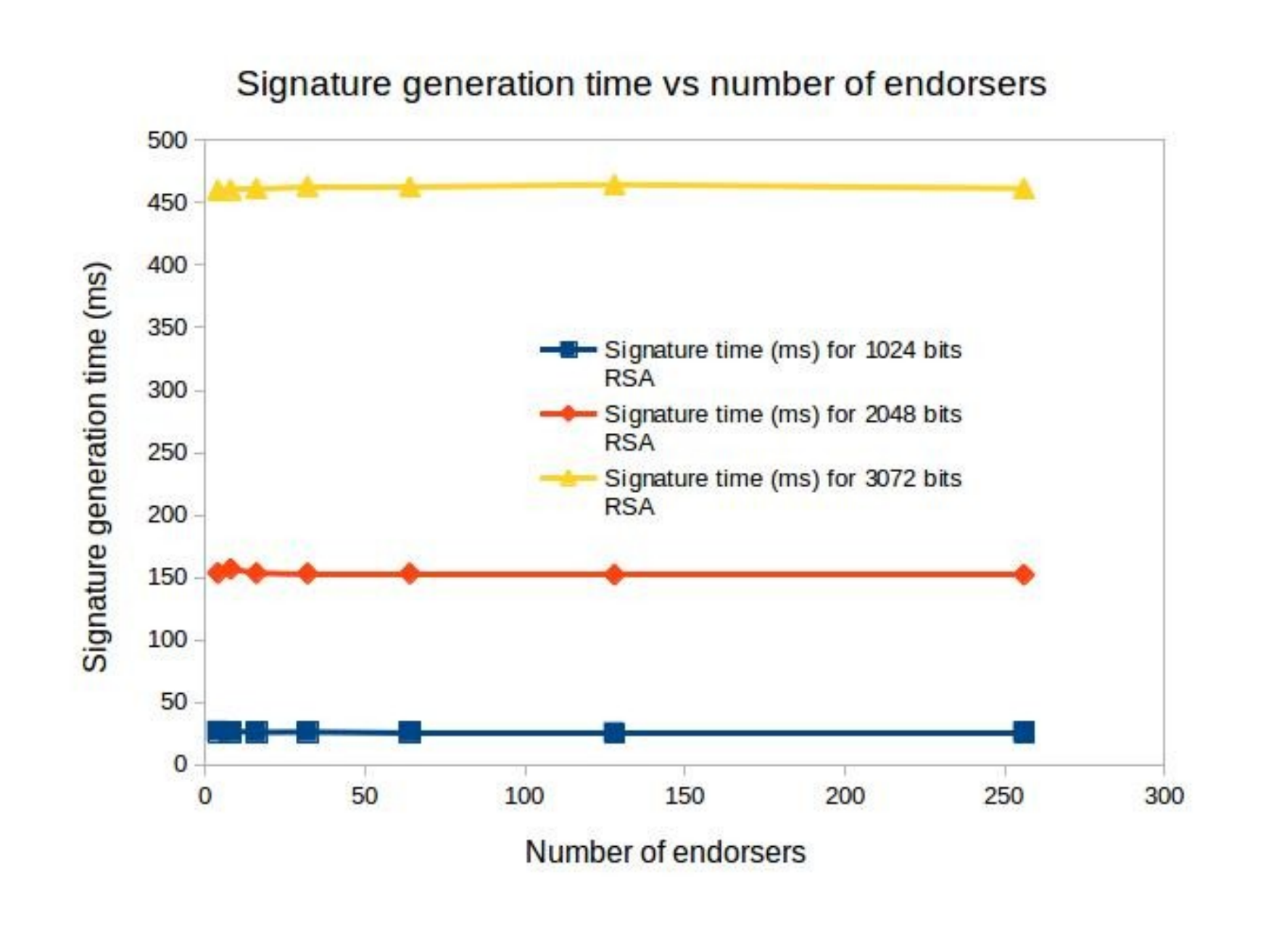}
  \caption{Signature Run time vs endorsement set size plot}
  \label{fig:graph}
\end{figure}

\begin{figure}[H]
\centering
  \includegraphics[scale=0.6]{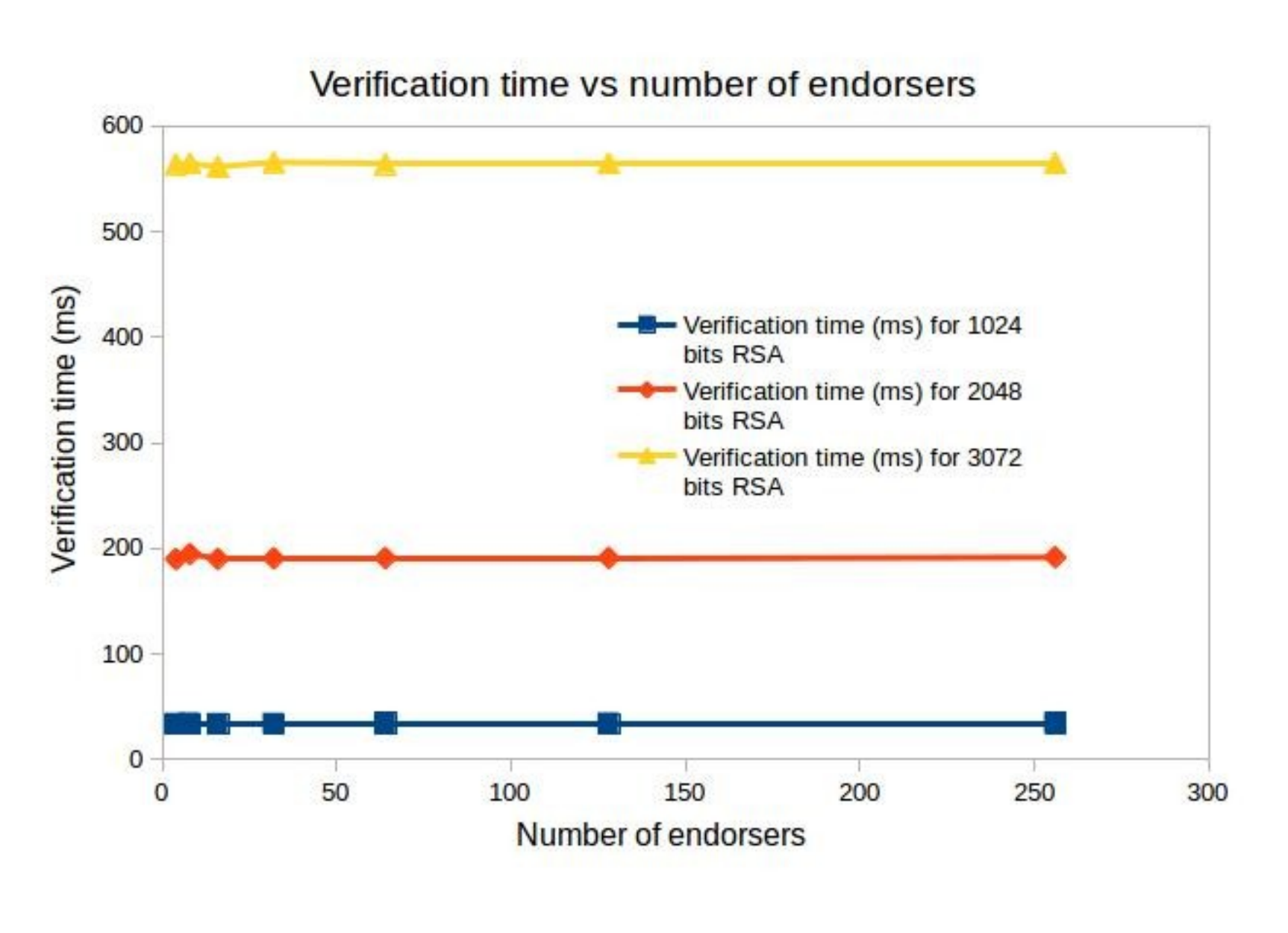}
  \caption{Verification Run time vs endorsement set size plot}
  \label{fig:graph1}
\end{figure}

Time taken by the signature generation and verification algorithm with respect to number of endorsers has been plotted in Fig. \ref{fig:graph} and Fig. \ref{fig:graph1} respectively. The signature generation time and verification time remains constant for a fixed value of RSA modulus size. On varying the $\lambda$ value, increase in computation time has been observed for both the algorithms. The Golang code for signature generation  of \textit{1-out-of-n} endorsement policy is available in \cite{Code}. \textcolor{black}{ The performance of the scheme measured through experiments also supports the theoretic analysis in Table \ref{asym}. For threshold value $t, 1 \leq t \leq n$, the signature generation time and tag generation time is same as that recorded for one signer since every signer will generate signature in parallel. The verification time becomes $t$ times that of the time taken to verify a single signature. This is because each verifier verifies $t$ such valid signatures sequentially.}
\begin{table}[H]
\caption{Signature generation time for FCsLRS vs number of participants}  
  \centering
  \scalebox{0.9}[0.9]{
  \begin{tabular}{|c |c | c |c|}
   \hline
  Endorsement set &Signature generation  &Signature generation    &Signature generation\\
  size           &  time(ms) for 1024 bit    &time(ms) for 2048 bit  &time(ms) for 3072 bit\\
                  & RSA modulus             & RSA modulus &RSA modulus \\
 \hline

 4	&26.3810697	&153.494131	&460.1620130\\
8	&25.9974438	&156.6600916 &460.5884873 \\
16	&25.9047032	&153.3631163	&461.7230385 \\
32	&26.0976248	&153.1091731	&463.5039483 \\
64	&25.9170626	&153.0620514	&462.9505392 \\
128	&25.187382	&152.0527724	&464.6665291 \\
256	&25.6127477	&152.1151231	&461.6722804 \\
\hline
%
%
%
\end{tabular}
}
\label{tab2}
\end{table}
\begin{table}[H]
\caption{Verification time for FCsLRS vs number of participants}  
  \centering
  \scalebox{0.9}[0.9]{
  \begin{tabular}{|c |c | c |c|}
   \hline
  Endorsement set &Verification time(ms)   &Verification time(ms)    &Verification time(ms)\\
  size           &  for 1024 bit    &for 2048 bit  &for 3072 bit\\
                  & RSA modulus             & RSA modulus &RSA modulus \\
 \hline
4	&33.8444186	&189.9202852	&563.9633690 \\
8	&34.1522705	&194.7279413	&564.7610533 \\
16	&33.7478959	&190.1404563	&561.8273189 \\
32	&33.8859284	&190.7751803	&565.7962644 \\
64	&34.4397107	&190.8586968	&564.2824740 \\
128	&34.1952231	&191.0990971	&564.6969238 \\
256	&34.3507846	&191.6187186	&565.3521524 \\
\hline
\end{tabular}
}
\label{tab3}
\end{table}

\section{Integration of Constant-Sized Linkable Ring Signature module in Hyperledger Fabric}
\label{integrate}
\begin{figure}
\centering
\subfloat[PROPOSE message format]{\includegraphics[width=0.5\textwidth]{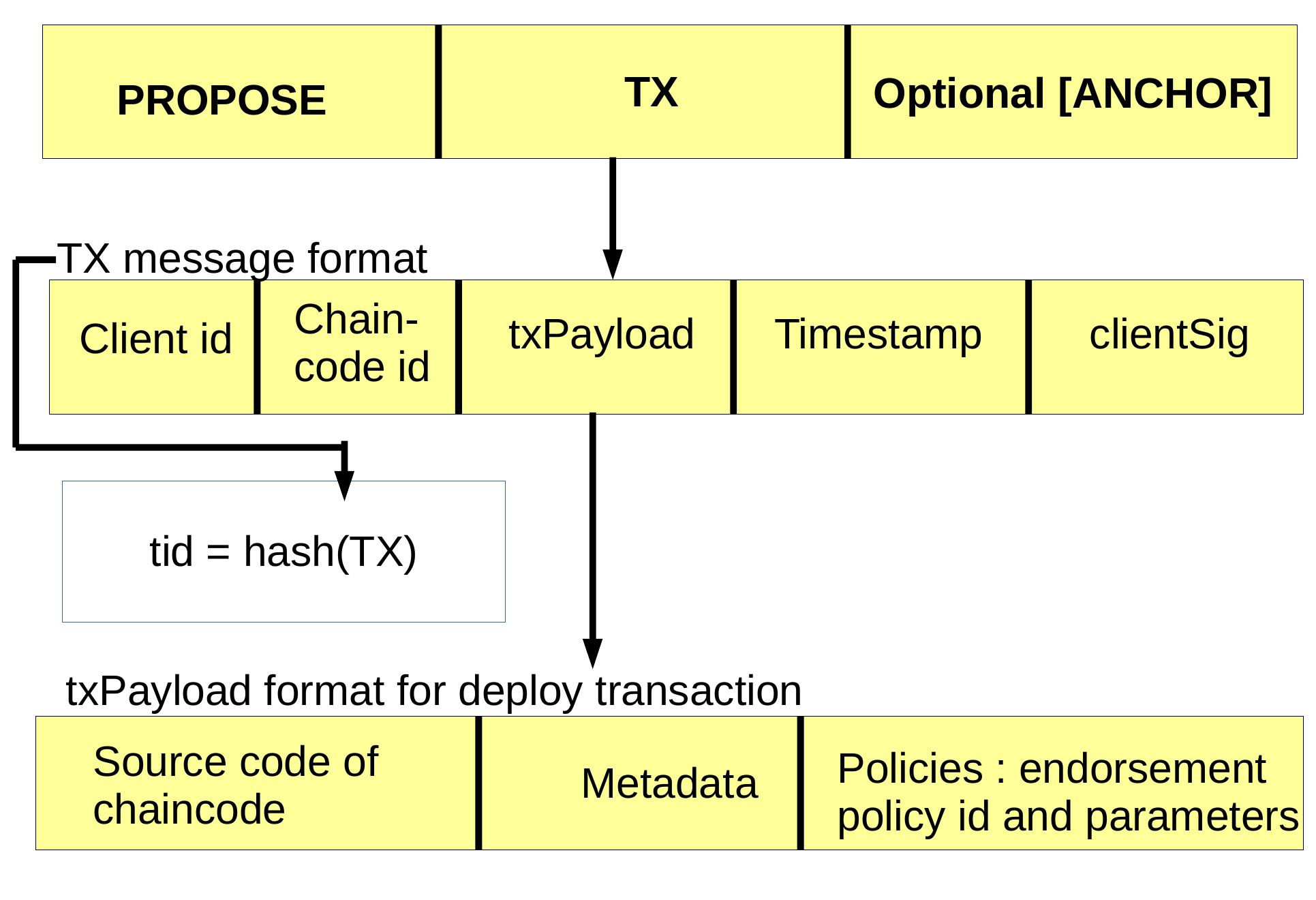}\label{prp1}}
  \hfill
  \subfloat[PROPOSAL RESPONSE message format]{\includegraphics[width=0.45\textwidth]{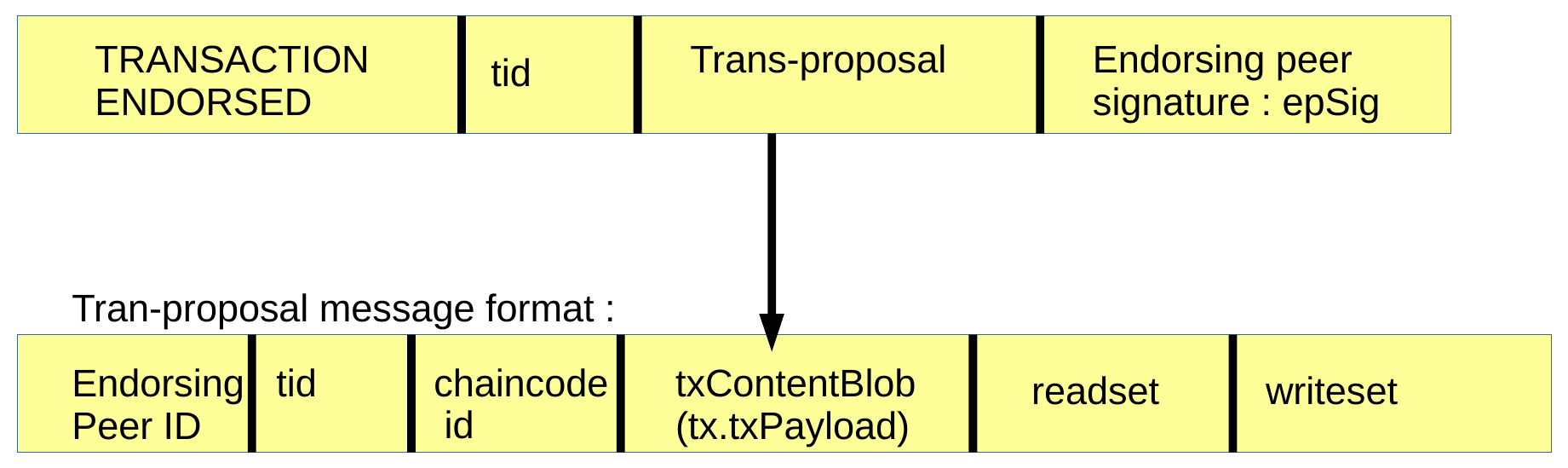}\label{prp2}}
 \caption{}
\end{figure}

In current workflow of transaction (\cite{Hyperledger}), to invoke a transaction, the client sends a ``\texttt{PROPOSE} message'' to the endorsers mentioned in the endorsement policy, provided endorsers are also part of the given \textit{chaincodeID}. 
The format of a \texttt{PROPOSE message} is \textit{$\langle$PROPOSE,tx,[anchor]$\rangle$} as shown in Fig. \ref{prp1}.

The endorser with identity \textit{epID}, on receipt of the \texttt{PROPOSE message} message,verifies the client’s signature denoted by \textit{clientSig}. It then executes the transaction (txPayload) by forwarding internally the \texttt{tran-proposal} to the part of its logic that endorses a transaction. Currently, the endorsing logic by default accepts the \texttt{tran-proposal} and signs the \texttt{tran-proposal}. However, one can change the endorsing logic as per the requirement to reach a decision whether to endorse a transaction or not. After endorsing the transaction, the peer sends a \texttt{\textcolor{black}{PROPOSAL-RESPONSE}} packet containing the message - \textit{$\langle$TRANSACTION-ENDORSED, tid, \texttt{tran-proposal},epSig$\rangle$} (as shown in Fig. \ref{prp2}) to the submitting client, where: \texttt{tran-proposal} := \textit{(epID,tid,chaincodeID,txContentBlob, readset,writeset)}. txContentBlob denotes chaincode/transaction specific information and \textit{epSig} denotes the endorsing peer’s signature on \texttt{tran-proposal}. If in case the endorsing logic refrains from endorsing the transaction, an endorser may send a negative acknowledgment to the submitting client stating its decision of rejecting the transaction\cite{Hyperledger}.

\begin{figure}[!ht]
\centering
\includegraphics[scale=0.5]{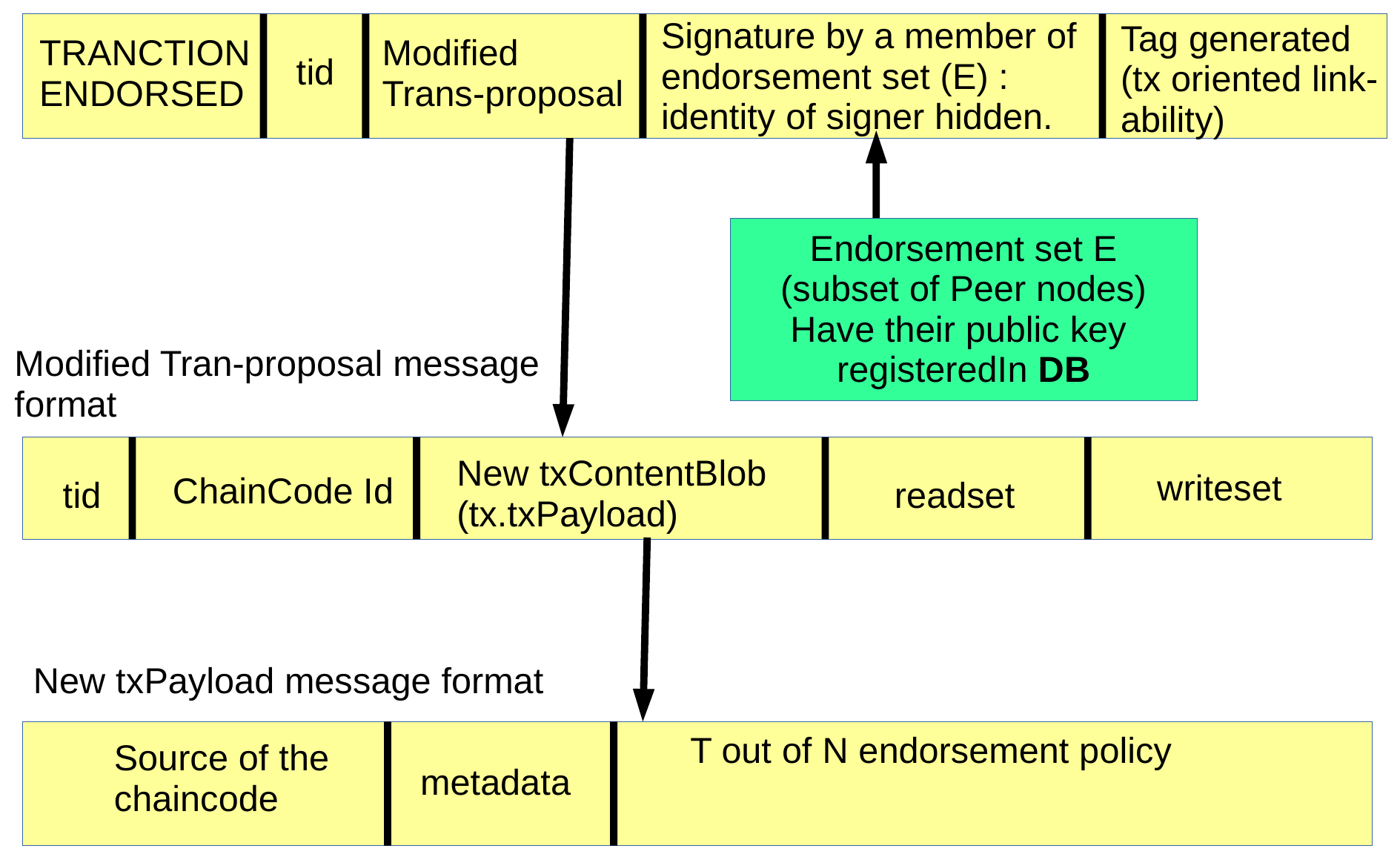}
\caption{Modified PROPOSE message format}
\label{prp3}
 \end{figure}
 
\begin{figure}[!ht]
\centering
\includegraphics[width=12cm,height=10cm,keepaspectratio]{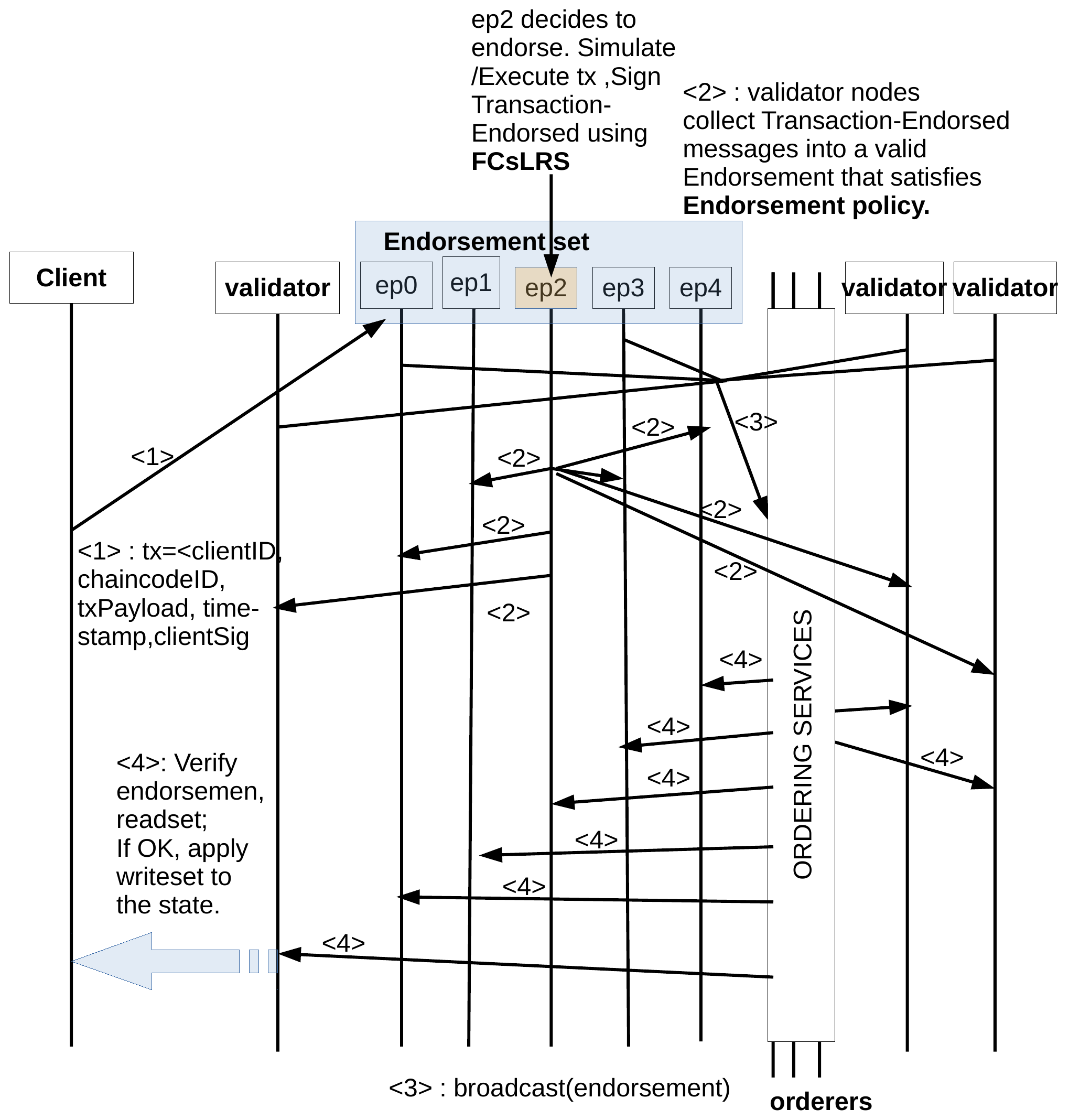}
\caption{Modified transaction flow diagram : 1-out-of-N endorsement policy}
\label{prp4}
 \end{figure}
To integrate the \textit{Constant-Sized linkable ring signature} module, we first have to change the \textbf{PROPOSAL RESPONSE} format (as shown in Fig. \ref{prp3}). In the source code, under \textit{hyperledger/fabric/protos/peer/  proposal\_response.proto} in the structure \textbf{ProposalResponse}, add a field called as \textbf{Tag} which will enable \textit{Transaction-oriented} linkability. The structure \textbf{Endorsement} must be changed by deletion of the field \textbf{endorser} (data type bytes[ ]) which reveals the endorsing peer ID. 
Create a \textbf{FCsLRS}\footnote{Fabric's Constant-Sized Linkable Ring Signature} package under \textit{hyperledger/fabric/bccsp} which can be used by the signer to sign the message. In the file \textit{hyperledger/fabric/msp/identities.go}, delete the field \textbf{identity} in the structure \textbf{signingidentity}. The \textbf{Verify} function will just check the validity of the signature corresponding to a message. Instead of checking the identity of signer, it will verify whether signer can give a \textit{Signature based on Proof of Knowledge} of the secret to prove its membership to the \textit{Endorsement set} $\mathcal{E}$. 

The \textit{Transaction flow diagram} for \textit{1-out-of-N'} endorsement policy is given in Fig. \ref{prp4}. When \textit{endorsing peer ep2} decides to endorse the transaction, it constructs a ring signature using public key of members in the \textit{Endorsement set} and the \texttt{PROPOSAL RESPONSE} format being same as that shown in Fig. \ref{prp3}. This is broadcasted to all the \textit{Peer} nodes ($\langle 2 \rangle$ of Fig. \ref{prp4}). If majority of the nodes reach a consensus on receipt of a valid endorsement, then one of the \textit{validator} node forwards it to the \textit{Ordering Service}($\langle 3 \rangle$ of Fig. \ref{prp4}). \textcolor{black}{Now ordering service broadcasts the packets to all peer nodes for \textit{validation} (step $\langle 4 \rangle$ of Fig. \ref{prp4}). All peers deterministically validate the transactions in the same order by checking satisfaction of endorsement policy and version number of the keys present in the local key-value store.}

\section{Description of the Implementation }
\label{imp}
%
%
%

In this section, we give a high level description of the main methods for \textit{1-out-of-n} endorsement policy. Assuming that a signer $S$ wants to endorse a transaction with transaction id as $tid=hash(transaction \ payload)$ and transaction payload denoted by $m \in \mathcal{M}$, where $\mathcal{M} \in {0,1}^*, S \in \mathcal{E}$, $\mathcal{E}$ is the endorsement set, secret key is $sk_S=(p_\pi,q_\pi)$ and public key of $S$ is $pk_S=2p_\pi.q_\pi+1$. Since we consider the case of just one endorser, we eliminate the code for check of linkability match as of now. But when it is integrated, check for linkability must be added for each transaction.
\begin{itemize}
\item[1.] \textbf{Initialization.} This step involves generation of the RSA Modulus integer $N$ of size $\lambda$ bits. This step is executed by Fabric CA which generates  the values by taking the security parameters as its input. To find a generator of $QR(N)$, we use the following lemma (\cite{micciancio2005rsa}) :
\begin{lemma}
\label{qr}
Let $N = p.q$ be the product of two distinct safe primes, and $u \in QR(N)$ a quadratic residue. Then $u$ is a generator for $QR(N)$ if and only if $gcd(u - 1, N ) = 1$.
\end{lemma}
\begin{proc}
    \SetKwInOut{Input}{Input}
    \SetKwInOut{Output}{Output}

     \Input{ $\lambda$ }
     \Output{Public parameters : $N,g,h,t,y,s,\zeta $}
    \caption{Initializations}
        \label{algo:initia}
     \begin{enumerate}
    \item Generate 2 safe primes $p,q : p=2p'+1, q=2q'+1, |p|=|q|=\frac{\lambda}{2}$.
    \item Find $N=p.q$.
    \item Find a generator of the group $QR(N)$ using Lemma \ref{qr}. Let that be $u$.
    \item $u$ generates $g,h,t,y,s,\zeta$ using some random discrete logarithm value $rd_i, 1\leq i \leq 6, 2 \leq rd_i  \leq |QR(N)|-1 $ where $|QR(N)|=p'.q'$.
     
	\end{enumerate}        
\end{proc}
\item[2.] \textbf{Key Generation.} Given an input $n$, which is the number of endorsers, each of the endorsers generate their own public key and private key pairs independently. (It only proves using zero-knowledge to Fabric CA about the correctness of the public key generated \footnote{In our implementation, since we have developed the signature scheme as an independent module without considering any Public Key Infrastructure, so for the ease of implementation we have assumed the public keys generated by each endorser is correct.}). Upon key generation , these values are made available in the public database $\mathcal{DB}$. The procedure mentioned below must be run parallely for each endorser present in endorsement set $\mathcal{E}$.   

\begin{proc}
    \SetKwInOut{Input}{Input}
    \SetKwInOut{Output}{Output}

     \Input{ $\lambda, l , \mu : \lambda > l - 2 , \frac{l}{2} > \mu + 1$, where $\mathcal{E}$ is the endorsement set }
      
     \Output{Public Key : $pk_i$, Secret Key : $sk_i$}
    \caption{Key Generation for endorser $E_i$}
        \label{algo:keygen}
     \begin{enumerate}
     \item Generate 2 prime $p,q, p \neq q : q \in (2^\frac{l}{2}-2^\mu+1,2^\frac{l}{2}+2^\mu-1)$. $sk_i=(p,q)$.
    \item Generate $pk_i : pk_i=2p.q+1$.
    \item Send $pk_i$ to database $\mathcal{DB}$ .
        
	\end{enumerate}        
\end{proc}
\item[3.]\textbf{Public Key accumulation}. Fabric CA uses its accumulator with one-way domain to generate an accumulated value of all the public keys in $\mathcal{DB}$, each having valid enrolment certificate.   

\begin{proc}
    \SetKwInOut{Input}{Input}
    \SetKwInOut{Output}{Output}

     \Input{ all $pk$'s in database $\mathcal{DB}$, generator $u, \langle u \rangle = QR(N)$ }
      
     \Output{Accumulated value : $v$}
    \caption{Accumulated value computation}
        \label{algo:acc}
     $v \leftarrow u$ \\
     \For{ $pk_i \in \mathcal{DB}$}
     {
         $v \leftarrow v^{pk_i} \mod N$
     }
\end{proc}

\item[4.]\textbf{Witness Generation for Signer S}. Signer $\mathcal{S}$ can generate the witness $w$ using values of all public keys forming the ring except its own public key.
\begin{proc}
    \SetKwInOut{Input}{Input}
    \SetKwInOut{Output}{Output}

     \Input{ all $pk$'s in database $\mathcal{DB}$, generator $u, \langle u \rangle = QR(N)$, Signer $S$ public key : $pk_S$ }
      
     \Output{witness value : $w$}
    \caption{Witness value for signer $S$}
        \label{algo:wit}
     $w \leftarrow u$ \\
     \For{ $pk_i \in \mathcal{DB} : pk_i \neq pk_S$ }
     {
         $w \leftarrow w^{pk_i} \mod N$
     }
\end{proc}
\item[5.]\textbf{Tag Generation}. To generate the tag, the signer $S$ needs to compute $g_tid$ from $g$ given the transaction id \textit{tid}.
\begin{proc}
    \SetKwInOut{Input}{Input}
    \SetKwInOut{Output}{Output}

     \Input{ Signer $S$ secret key : $sk_S=(p_\pi,q_\pi)$, transaction id : tid }
      
     \Output{tag value : $\tilde{y}$}
    \caption{Tag generation for signer $S$}
        \label{algo:tag}
        $x \leftarrow \tilde{H}(tid)$ \\
        $g_{tid} \leftarrow g^x \mod N$\\
        $\tilde{y}\leftarrow g_{tid}^{p_\pi+q_\pi} \mod N$\\
%
\end{proc}

\item[6.] \textbf{Computation of public values for Signature based on Proof of Knowledge Construction}. Signer $S$ computes public values $T_1,T_2,T_3,T_4,T_5$ where

 $T_1=g_{tid}^r \mod N , T_2= (h^r \zeta^{pk_S+r}) \mod N, T_3= (s^r g_{tid}^{q_\pi}) \mod N, T_4=(w.y^r)\mod N, T_5=(t^r g_{tid}^{2p_\pi})\mod N$.
\begin{proc}
    \SetKwInOut{Input}{Input}
    \SetKwInOut{Output}{Output}

     \Input{ Signer $S$ secret key : $sk_S=(p_\pi,q_\pi)$, public key $pk_S$, $g_tid$, witness value : $w$ }
      
     \Output{Public values : $T_1,T_2,T_3,T_4,T_5$}
    \caption{Public value generation by signer $S$}
        \label{algo:pv} 
        $r \xleftarrow{R} \mathbb{Z}_{N/4} $ \\
        Compute $T_1,T_2,T_3,T_4,T_5$ as per equations mentioned above.\\
%
\end{proc}
\item[7.]\textbf{Signature Generation}. Signer generates the challenge value which can be generated again at the verifier side as well. This is the standard \textit{Fiat-Shamir Transformation} which has been used. Send all these value (mentioned in the output of \textit{Signature Algorithm} along with tag $\tilde{y}$ to verifier $v \in \mathcal{V}$.
\begin{proc}
    \SetKwInOut{Input}{Input}
    \SetKwInOut{Output}{Output}

     \Input{$r$,Signer $S$ secret key : $sk_S=(p_\pi,q_\pi)$, public key $pk_S$, message $m \in \mathcal{M}$}
     \Output{ $u_1,u_2,\ldots,u_9,\tilde{\alpha_1},\tilde{\alpha_2},\ldots,\tilde{\alpha_5}$ }
    \caption{Signature}
        \label{algo:sign}
     \begin{enumerate}
     \item Generate $\alpha_i, 1 \leq i \leq 3 : 0 < \alpha_i < N/4 -1 $.
     \item Compute $u_1,u_2,\ldots,u_9$ as per Eq. \ref{sign}. 
     \item Computes the challenge value $c=H_1(m,u_1,u_2,\ldots,u_9)$, where $H_1$ is a random oracle.
     \item Using $c$, compute $\tilde{\alpha_1},\ldots,\tilde{\alpha_5}$ as per Eq. \ref{comm}.
   
	\end{enumerate}        
\end{proc}
\item[8.] \textbf{Verification Algorithm}. Verifier $v$ computes $c$ using the values sent to it by signer $S$. Using equations under Eq. \ref{ver} it's going to check whether the \textit{Signature based on Proof of Knowledge} construction is correct or not.
\end{itemize}

\begin{figure}
The main methods described in section \ref{imp} is illustrated here.
\centering
  \includegraphics[scale=0.7]{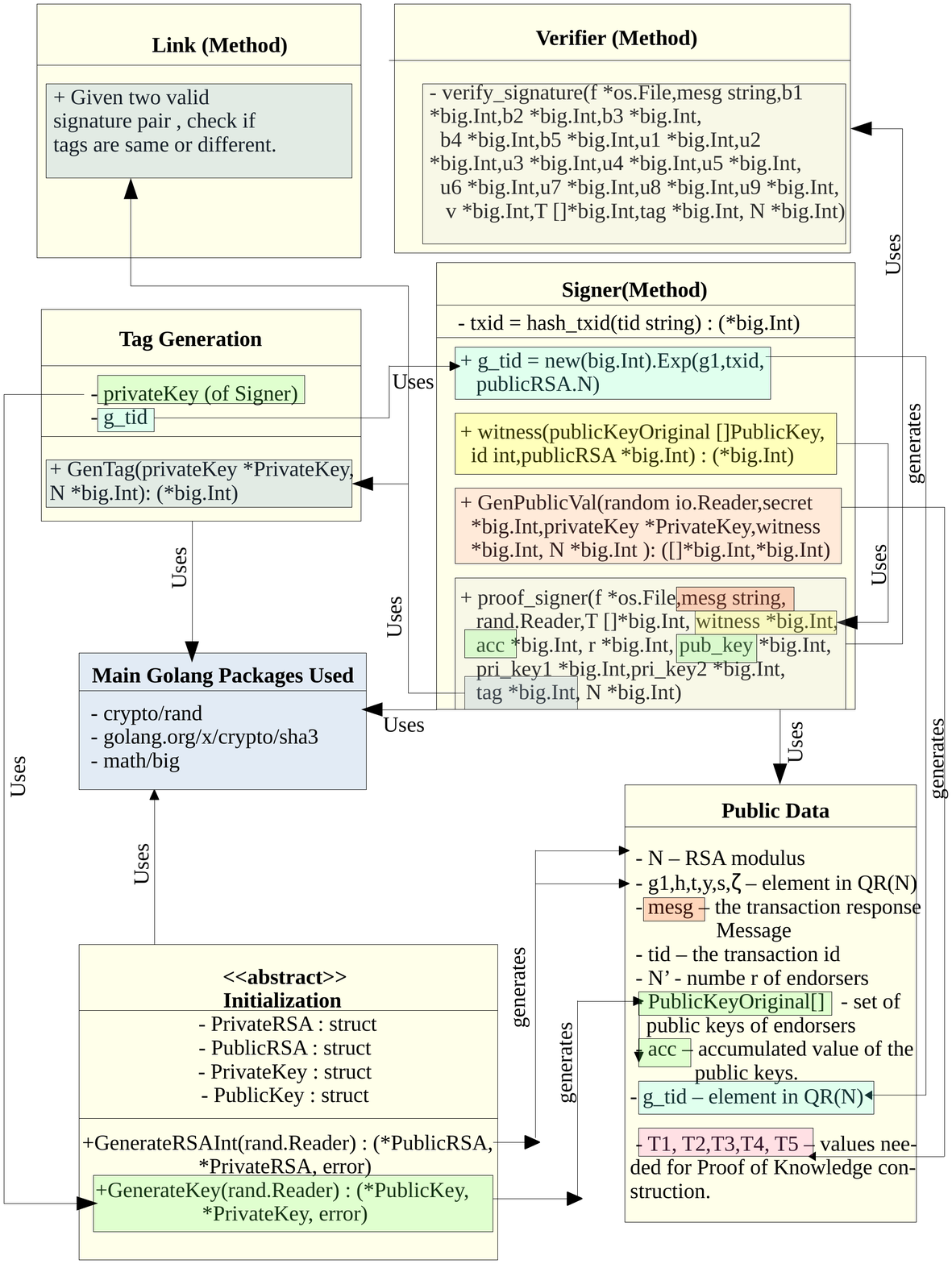}
  \caption{Class Diagram}
  \label{fig:img1}

\end{figure}

\newpage
\section{Conclusion and Future Work}
\label{conc}
\textcolor{black}{Our proposed ``Anonymous Endorsement System'' addresses the problem of biased voting by endorsers in Hyperledger Fabric. We have given a construction of new constant sized linkable ring signature scheme, FCsLRS which hides identity of each endorser involved in endorsement. The signature scheme can be independently used in e-voting, using voter identity linkability where a single person cannot vote more than once. Similarly, in public pages of social media, we can use this endorsement system for posts or digital content that needs to be validated by moderators to prevent false propaganda leading to a negative impact in the society. The moderators with certain reputation level may form the ring members and whoever amongst them supports the view can endorse it. If the post acquires adequate number of votes it will get displayed publicly. } 

\textcolor{black}{Since it is an RSA based scheme, it will require inclusion of RSA module for incorporating this module in Hyperledger Fabric as currently ECDSA (Elliptic Curve Digital Signature Algorithm) is supported. Inclusion of this module will allow us evaluate the performance in the actual setting.} 

\textcolor{black}{The major limitation of the scheme is the verifiers are required to count individual valid ring signature and check if the aggregate is above the threshold in order to implement threshold endorsement policy. This increases the verification time depending on the threshold value.} We would like to replace it with threshold signature scheme which can guarantee the same level of anonymity as it is offered now by the proposed system. 

In our future work, we aim to provide a construction of short ring signature scheme, probably using pairing based cryptography. \textcolor{black}{Since we prove the security of our construction in \textbf{Random Oracle Model}, we would like to provide a construction in the standard model as well.} Apart from that, we would like to explore other permissioned blockchain systems and check whether the proposed scheme can be extended over there as well.

\section{Acknowledgement}
This work is partially supported by \textit{Cisco University Research Program Fund}, CyberGrants ID: $\#698039$ and \textit{Silicon Valley Community Foundation}. The authors would like to thank Chris Shenefiel, Samir Saklikar  and Anoop Nannra for their comments and suggestions.

%
%
%
%
%
%
%
%
%
%
%
%
%
%
%
%
%
%
%



\end{document}